\newif\ifpreprint

\preprintfalse 

\ifpreprint
\documentclass[journal=jctcce,manuscript=letter]{achemso}
\else
\documentclass[journal=jctcce,manuscript=letter,layout=twocolumn]{achemso}
\fi

\usepackage[T1]{fontenc} 

\usepackage{amsmath}
\usepackage{newtxtext,newtxmath}

\usepackage{graphicx}
\usepackage{dcolumn}
\usepackage{braket}
\usepackage{multirow}
\usepackage{threeparttable}
\usepackage{xspace}
\usepackage{verbatim}
\usepackage[version=4]{mhchem} 
\usepackage{comment}
\usepackage{color,soul}
\usepackage{siunitx}
\usepackage{physics}

\usepackage{mathtools}
\usepackage[dvipsnames]{xcolor}
\usepackage{xspace}
\usepackage{ifthen}

\usepackage{qcircuit}

\usepackage{graphicx,longtable,dcolumn,mhchem}
\usepackage{rotating,color}
\usepackage{lscape}
\usepackage{amsmath}
\usepackage{dsfont}
\usepackage{soul}
\usepackage{physics}
\newcolumntype{d}{D{.}{.}{-1}}

\newcommand{\ra}{\rightarrow}
\newcommand{\pis}{\pi^\star}
\newcommand{\npi}{n\ra\pis}
\newcommand{\ppi}{\pi\ra\pis}

\newcommand{\Pop}{6-31+G(d)}
\newcommand{\AVDZ}{{aug}-cc-pVDZ}
\newcommand{\AVTZ}{{aug}-cc-pVTZ}

\usepackage[normalem]{ulem}

\usepackage[colorlinks = true,
            linkcolor = blue,
            urlcolor  = blue,
            citecolor = blue,
            anchorcolor = blue]{hyperref}
\definecolor{goodorange}{RGB}{225,125,0}
\definecolor{goodgreen}{RGB}{5,130,5}
\definecolor{goodred}{RGB}{220,50,25}
\definecolor{goodblue}{RGB}{30,144,255}

\newcommand{\note}[2]{
\ifthenelse{\equal{#1}{F}}{
\colorbox{goodorange}{\textcolor{white}{\footnotesize \fontfamily{phv}\selectfont #1}}
    \textcolor{goodorange}{{\footnotesize \fontfamily{phv}\selectfont #2}}\xspace
}{}
\ifthenelse{\equal{#1}{R}}{
\colorbox{goodred}{\textcolor{white}{\footnotesize \fontfamily{phv}\selectfont #1}}
    \textcolor{goodred}{{\footnotesize \fontfamily{phv}\selectfont #2}}\xspace
}{}
\ifthenelse{\equal{#1}{N}}{
\colorbox{goodgreen}{\textcolor{white}{\footnotesize \fontfamily{phv}\selectfont #1}}
    \textcolor{goodgreen}{{\footnotesize \fontfamily{phv}\selectfont #2}}\xspace
}{}
\ifthenelse{\equal{#1}{M}}{
\colorbox{goodblue}{\textcolor{white}{\footnotesize \fontfamily{phv}\selectfont #1}}
    \textcolor{goodblue}{{\footnotesize \fontfamily{phv}\selectfont #2}}\xspace
}{}
}

\usepackage{titlesec}

\usepackage[fontsize=11pt]{scrextend}
\titleformat{\section}
{\normalfont\sffamily\bfseries\color{Blue}}
{\thesection.}{0.25em}{\uppercase}

\titleformat{\subsection}
{\normalfont\sffamily\bfseries}
{\thesubsection}{0.25em}{}

\titleformat{\subsubsection}
{\normalfont\sffamily}
{\thesubsubsection}{0.25em}{}

\titleformat{\suppinfo}
{\normalfont\sffamily\bfseries}
{\thesubsection}{0.25em}{}

\titlespacing*{\section}{0pt}{0.5\baselineskip}{0.01\baselineskip}
\titlespacing*{\subsection}{0pt}{0.125\baselineskip}{0.01\baselineskip}
\titlespacing*{\subsubsection}{0pt}{0.125\baselineskip}{0.01\baselineskip}

\newcommand{\CEISAM}{Nantes Universit\'e, CNRS,  CEISAM UMR 6230, F-44000 Nantes, France}
\newcommand{\UOP}{Dipartimento di Chimica e Chimica Industriale, University of Pisa, Via Moruzzi 3, 56124 Pisa, Italy}
\newcommand{\CEA}{Universit\'e Grenoble Alpes, CEA, IRIG-MEM-L Sim, 38054 Grenoble, France}
\newcommand{\UGA}{Universit\'e Grenoble Alpes, CNRS, Institut NEEL, F-38042 Grenoble, France}
\newcommand{\LCPQ}{Laboratoire de Chimie et Physique Quantiques, Universit\'e de Toulouse, CNRS, UPS, France}
\newcommand{\IUF}{Institut Universitaire de France (IUF), F-75005 Paris, France}

\author{Iryna Knysh}
	\affiliation[UN, Nantes]{\CEISAM}    
\author{Filippo Lipparini}
	\affiliation[UP, Pisa]{\UOP}
\author{Aymeric Blondel}
	\affiliation[UN, Nantes]{\CEISAM}  
\author{Ivan Duchemin}
	\affiliation[CEA, Grenoble]{\CEA}
\author{Xavier Blase}
	\affiliation[NEEL, Grenoble]{\UGA}
\author{Pierre-Fran{\c c}ois Loos}
	\email{loos@irsamc.ups-tlse.fr}
	\affiliation[LCPQ, Toulouse]{\LCPQ}
\author{Denis Jacquemin}
	\email{Denis.Jacquemin@univ-nantes.fr}
	\affiliation[UN, Nantes]{\CEISAM}    
	\alsoaffiliation[IUF, Paris]{\IUF}

\let\oldmaketitle\maketitle
\let\maketitle\relax
     \title{Reference CC3 Excitation Energies for Organic Chromophores: Benchmarking TD-DFT, BSE/$GW$ and Wave Function Methods
}
\date{\today}

\begin{tocentry}
	\centering
	\includegraphics[width=.72\textwidth]{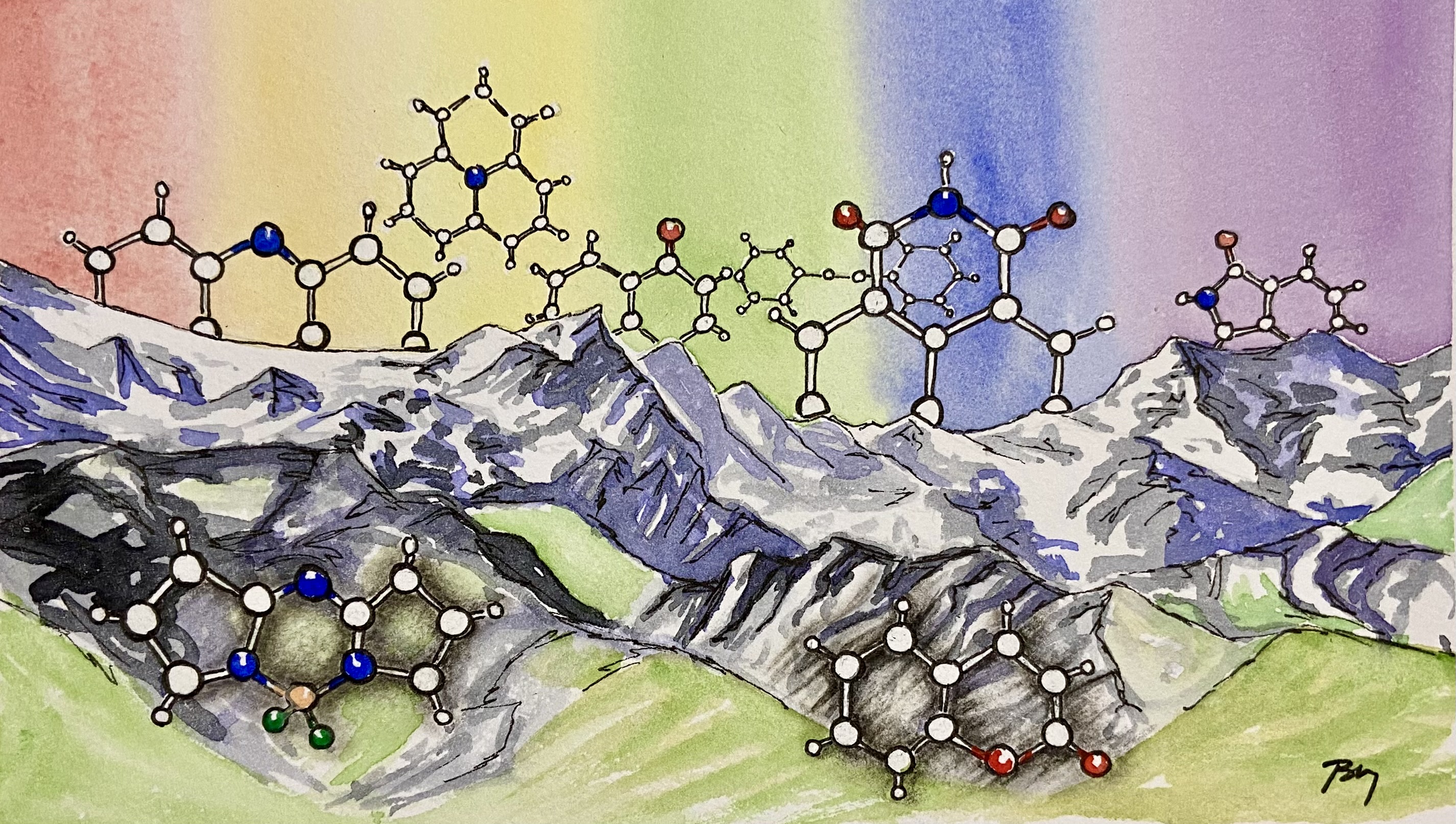}
\end{tocentry}

\begin{document}

\ifpreprint
\else
\twocolumn[
\begin{@twocolumnfalse}
\fi
\oldmaketitle

\begin{abstract}
To expand the QUEST database of highly-accurate vertical transition energies, we consider a series of large organic chromogens ubiquitous in dye chemistry, such as anthraquinone, azobenzene, BODIPY, 
and naphthalimide. We compute, at the CC3 level of theory, the singlet and triplet vertical transition energies associated with the low-lying excited states. This leads to a collection of more than 120 new 
highly-accurate excitation energies. For several singlet transitions, we have been able to determine CCSDT transition energies with a compact basis set, finding minimal deviations from the CC3 values. 
Subsequently, we employ these reference values to benchmark a series of lower-order wave function approaches, including the popular ADC(2) and CC2 schemes, as well as time-dependent density-functional 
theory (TD-DFT), both with and without applying the Tamm-Dancoff approximation (TDA). At the TD-DFT level, we evaluate a large panel of global, range-separated, local, and double hybrid functionals. 
Additionally, we assess the performance of the Bethe-Salpeter equation (BSE) formalism relying on both $G_0W_0$ and ev$GW$ quasiparticle energies evaluated from various starting points. It turns out 
that CC2 and ADC(2.5) are the most accurate models amongst those with respective $\mathcal{O}(N^5)$ and  $\mathcal{O}(N^6)$ scalings with system size. In contrast, CCSD does not outperform CC2. 
The best performing exchange-correlation functionals include BMK, M06-2X, M06-SX, CAM-B3LYP, $\omega$B97X-D, and LH20t, with average deviations of approximately 0.20 eV or slightly below. 
Errors on vertical excitation energies can be further reduced by considering double hybrids. Both SOS-$\omega$B88PP86 and SOS-$\omega$PBEPP86 exhibit particularly attractive performances with 
overall quality on par with CC2, whereas PBE0-DH and PBE-QIDH are only slightly less efficient.  BSE/ev$GW$ calculations based on Kohn-Sham starting points have been found to be particularly effective 
for singlet transitions, but much less for their triplet counterparts.
\end{abstract}

\ifpreprint
\else
\end{@twocolumnfalse}
]
\fi

\ifpreprint
\else
\small
\fi

\noindent

\section{Introduction}
\label{sec:Intro}

Organic dyes are everywhere, permeating countless aspects of our daily lives and playing pivotal roles in numerous applications. From the hues enhancing clothing and textiles to the intricate colors illuminating our 
electronic displays, organic dyes are indispensable to fashion, energy conversion, imagining, sensing, etc.  However, accurately simulating the colors produced by dyes and pigments poses a persistent challenge for 
theoretical chemists, largely owing to the remarkable sensitivity of the human eye, albeit confined to a limited region of the electromagnetic spectrum. Indeed, in the blue-green region, humans can distinguish color 
differences corresponding to changes in the absorption wavelength as tiny as $0.5$--$1.0$ nm  (approximately $0.005$ eV). Achieving this level of accuracy is currently beyond the capabilities of quantum many-body 
methods for molecules that absorb visible light.

\begin{figure*}[htp]
  \includegraphics[scale=0.52,viewport=1.5cm 2.5cm 29.0cm 18.5cm,clip]{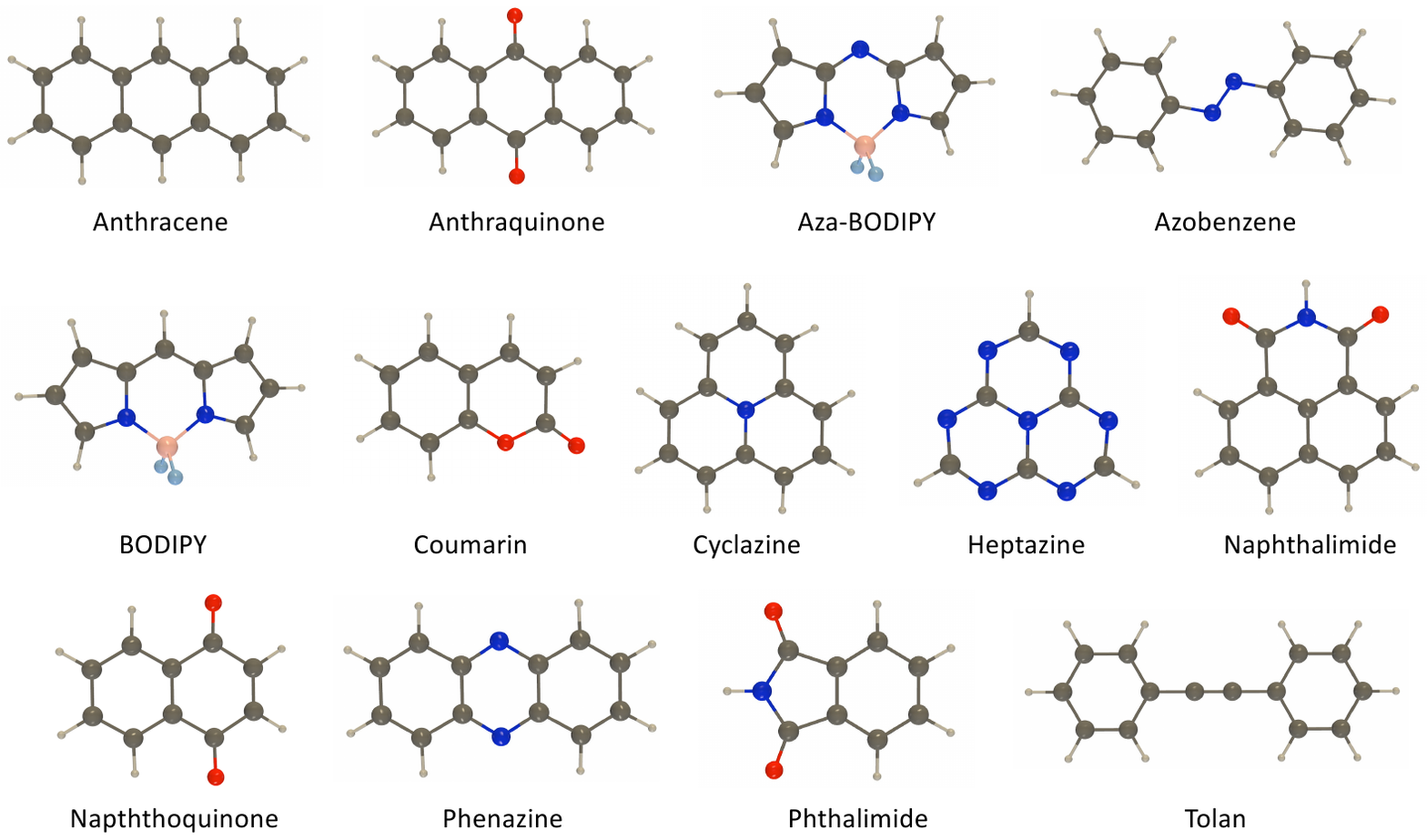}
  \caption{Representation of the systems investigated in the present study.}
  \label{Fig-1}
\end{figure*}

Despite this challenge, the quantum chemical modeling of absorption properties of organic dyes has been steadily advancing. A notable early contribution came from Adachi and Nakamura, who employed semi-empirical methods 
(CNDO/S and INDO/S) to investigate the excited-state (ES) properties of approximately 30 real-life dyes, drawn from popular chromophore families like azobenzene, anthraquinone, and hydrazone. \cite{Ada91}
In this work of 1991, the vertical transition energies (VTEs) were directly compared with the experimental wavelengths of maximal absorption ($\lambda_{\mathrm{max}}$) and the computed oscillator strengths ($f$) 
were correlated with the measured molar absorptivity coefficients ($\log \epsilon$), an approach that remains popular despite its intrinsic shortcomings. \cite{Lau14} Notably, the INDO/S method emerged as the most 
satisfactory, yielding a mean absolute error (MAE) of 0.48 eV and a linear correlation coefficient $R = 0.60$ when compared to experimental data. Clearly, these values fall short of enabling precise color prediction or the design 
of new dyes. 

A decade later, the same group proposed an analogous investigation for a similar set of transitions but relying  on time-dependent density-functional theory (TD-DFT) \cite{Cas95,Ulr12b} using two exchange-correlation 
functionals (XCFs): BPW91 and B3LYP.  \cite{Gui00} With the B3LYP global hybrid (GH), they reached a MAE of 0.19 eV, which they deemed \emph{``quite satisfactory''}. Later on, one of us applied TD-DFT on much 
larger and more diverse sets of organic dyes using an extended panel of XCFs. \cite{Jac08b,Jac09c} However, even with the most effective hybrid functional, PBE0, the MAE obtained remained similar at 0.22 eV.  

Two primary factors contribute to the magnitude of these errors.  
Firstly, the aforementioned works fail to consider vibronic couplings, which are crucial for achieving robust comparisons between experimental observations and theoretical predictions. 
\cite{Die04b,Imp07,Avi13,Cer16b,San16b}  Indeed, as demonstrated by Dierksen and Grimme two decades ago for a set of 43 transitions in $\pi$-conjugated organic derivatives, including vibronic couplings allows for
more meaningful comparisons between theory and experiment as band shapes are obtained. Nevertheless, the MAE on 0-0 energies remained essentially unchanged (0.18 eV with the BH{\&}HLYP functional). \cite{Die04b} 
Secondly, despite the elegance and potency of adiabatic TD-DFT, its inherent approximations do not allow to systematically achieve a high level of accuracy, particularly when relying on a single XCF, since ES of various natures 
ideally require different XCFs. For example, GHs are more suitable for valence transitions while charge-transfer (CT) excitations usually require range-separated hybrids (RSHs). \cite{Lau13} 

On the theoretical side, several methods are available to get more accurate VTEs. \cite{Loo20c} Firstly, one can further climb Jacob's ladder of DFT, opting for double hybrids (DHs) rather than GHs or RSHs. 
\cite{Goe09,Bre16,Sch17,Mes21} Secondly, another approach involves employing the Bethe-Salpeter equation (BSE) formalism \cite{Han79,Str88} of many-body perturbation theory \cite{Oni02,Mar16d} combined 
with the $GW$ method. \cite{Hed65,Ary98,Gol19,Mar23} The so-called BSE/$GW$ approach advantageously washes out the XCF dependency,  when a (partially) self-consistent $GW$ scheme is 
used. \cite{Bla18,Gui18b,Bla20}  {The diagonalization of the electron-hole BSE Hamiltonian in the (de)excitation space offers the same computational scaling as Casida's formulation 
of TD-DFT,} \cite{Cas95} {namely $\mathcal{O}(N^4)$ to get iteratively the lowest eigenvalues. Similarly, the preceding $GW$ calculations, required to get the quasiparticle energies and screened 
Coulomb potential that enter the BSE Hamiltonian, can be performed with $\mathcal{O}(N^4)$ scaling using resolution-of-the-identity (RI)  techniques.}
Thirdly, one can opt for correlated wave function approaches. Within this framework, the systematically improvable coupled-cluster 
(CC) route stands as a natural choice, offering a clear path for increases in accuracy with its CC2, \cite{Chr95,Hat00} CCSD,  \cite{Pur82,Scu87,Koc90b,Sta93,Sta93b} CC3, \cite{Chr95b,Koc95,Koc97} CCSDT,
\cite{Nog87,Scu88,Kuc01,Kow01,Kow01b} CC4,  \cite{Kal04,Kal05,Loo21b,Loo22} CCSDTQ, \cite{Kuc91,Hir04,Kal03,Kal04} and so forth declinations.  

This is the approach we primarily employed to build the QUEST database, \cite{Ver21} which currently includes over 500 VTEs considered \emph{chemically accurate}, that is, with a deviation below 0.05 eV from 
the exact result. QUEST relies on diffuse containing basis sets, and most results are, at least, of {\AVTZ} quality. While this value falls short of the human eye's ability in terms of accuracy, these reference 
VTEs are nonetheless useful for both methodological  benchmarking purposes \cite{Ver21} and for comparisons of various chemical strategies aiming at tailoring ES energies. In the initial release of the QUEST 
database, \cite{Ver21} the largest compounds considered were naphthalene and one of its tetra-aza derivatives. These two highly-symmetric bicyclic compounds contain 10 heavy (non-H) atoms. In 2021, we 
expanded this dataset by incorporating other, less symmetric bicyclic compounds, including real-life chromophores like diketopyrrolopyrrole, yet none larger than 10 non-H atoms. \cite{Loo21} A key finding from 
this 2021 work is the notable differences in the relative accuracies of the benchmarked wave function methods compared to those observed for smaller molecules. For instance, while CCSD provides very accurate 
VTE estimates for small molecules, its precision clearly diminishes for larger compounds. \cite{Ver21,Loo21} 

The present contribution aims at providing accurate VTEs for the series of real-life chromogens represented in Fig.~\ref{Fig-1}. We clearly draw inspiration from previous studies, \cite{Ada91,Gui00,Jac09c,Hoh19} with a focus on 
the most important classes of organic chromophores and fluorophores. \cite{Chr91,Zol03,Val04} Naturally, for molecules of such size, incorporating quadruple excitations at the CC level, as recently accomplished for five and 
six-membered rings, \cite{Loo22} or sufficiently converging selected configuration interaction calculations \cite{Hur73,Gar18,Gar19} is simply beyond reach. \cite{Eri20b,Loo20g,Eri21} However, obtaining CC3 values in a 
reasonably large basis set remains feasible. Moreover, it is even sometimes possible to perform CCSDT calculations, albeit in a relatively compact basis set. 

To the best of our knowledge, these are the first high-level calculations reported for a large panel of chromophores.  In fact, we are not aware of any previous CC3 calculation published for any of the compounds of 
Fig.~\ref{Fig-1}, except for CC3/{\AVDZ} values obtained for azobenzene, \cite{Hut20,Pau22} and our recent contribution on the lowest singlet and triplet ESs of cyclazine and heptazine computed at the CC3/{\AVTZ} 
level.\cite{Loo23a} 

There are nevertheless a few CC3 literature precedents that deserve to be highlighted, the first of which is the seminal series of works from Thiel and coworkers who obtained an impressive list of CC3/TZVP VTEs for 
organic compounds (the largest system being naphthalene). \cite{Sil08,Sil10b}  There are also more specific works using CC3 on rather large molecules. For example, Zuev \emph{et al.}~investigated two isomers of 
$p$-coumaric acid (12 non-H atoms) at the CC3/6-31+G(d,p) level, \cite{Zue11} Kannar and Szalay studied numerous ESs of the DNA bases (largest one being composed of 10 non-H atoms) at CC3/TZVP 
level, \cite{Kan14b} Hutcheson and coworkers modeled the two lowest ESs of azobenzene (14 non-H atoms) at the CC3/{\AVDZ} level, Cuzzocrea \emph{et al.}~used CC3/{\AVDZ} to tackle increasingly long cyanine 
chains up to 15 non-H atoms, \cite{Cuz22} Gui and coworkers obtained CC3/TZVP estimates for a series of push-pull systems (up to 11 non-H atoms), \cite{Gui18b} and we obtained CC3/cc-pVTZ values for 28 CT 
ESs (as large as 12 non-H atoms). \cite{Loo21a}  In these two latter cases, further basis set corrections were applied to account for the lack of diffuse basis functions. 

Even larger systems can be considered if one relies on local correlation methods.  
Notably, H\"attig's group performed a CC3/{\AVTZ} calculation of the lowest ES of berenil (29 non-H atoms) using their pair-natural-orbital implementation, \cite{Fra20} whereas Koch's group determined 
multi-level/natural-transition-orbital CC3/{\AVDZ} valence VTEs on azobenzene (14 non-H atoms), \cite{Pau21,Pau22} and the core excitations of betaine30 (43 non-H atoms). At the CCSD level of theory, the 
number of studies is too large to be listed. Yet, we wish to highlight the 2019 work of Martinez's group. \cite{Hoh19} Using a rank-reduced version of CCSD, they tackled 19 ESs in 8 real-life dyes. \cite{Hoh19} 
This represents the closest attempt at using high-level CC methods for estimating VTEs in large dyes. In addition, many works rely on CC2 or its algebraic diagrammatic construction counterpart, ADC(2), for estimating 
VTEs in photoactive systems (see, for example, Refs.~\citenum{Goe10a,Sen11,Win13,Jac15b,Oru16}).

This paper is organized as follows. In Section \ref{sec:Comp}, we detail the protocols used to perform our calculations. In Section \ref{sec:Res}, we first present the obtained reference values and comment on their
nature and accuracy. We then provide comparisons with selected literature data. Next, we turn to the benchmark of wave function-based, density-based, and one-body Green's function-based approaches before 
closing the result section with further discussion about the change of reference values and a specific challenging case.  Section \ref{sec:CCl} summarizes our main findings.

\section{Computational details}
\label{sec:Comp}

\subsection{Generalities}

All calculations presented in this study are conducted, except when noted, under the frozen-core approximation and typically use the default convergence thresholds of the chosen electronic structure softwares. 
We underline that distinct symmetry conventions concerning  molecular orientations in various codes may lead to distinct state labeling for molecules belonging to the $C_{2v}$ ($B_1 \leftrightarrow B_2$) and 
$D_{2h}$ ($B_{1x} \leftrightarrow B_{2x} \leftrightarrow B_{3x}$, with $x = g$ or $u$) point groups. To address this potential difficulty, we provide in the Supporting Information (SI) various information regarding 
each state (see below) enabling quick identification of the ESs of interest. It is worth mentioning that, in the subsequent discussion, we do not specify the equation-of-motion (EOM) or linear-response (LR) prefixes 
for the CC methods, as both formalisms yield identical VTEs. The CCSD/{\AVTZ} oscillator strengths listed in the SI have been determined with LR-CCSD.

\subsection{Geometries}

The ground-state structures of all compounds were optimized at the CCSD(T)/cc-pVTZ level using \textsc{cfour} 2.1,  \cite{cfour,Mat20} building $Z$-matrices to enforce the relevant point group symmetry. For both 
aza-BODIPY and BODIPY,  the $C_{2v}$ point group was selected for obvious computational reasons, though the fluoroborate moiety should normally be slightly out-of-plane leading to $C_s$ ground-state structures.  The optimal 
structures of all compounds are shown in Figure \ref{Fig-1} and Cartesian coordinates are provided in the SI.

\subsection{Reference Excitation Energies}

For a given molecule, we typically start our exploration by performing a CCSD/{\AVTZ} calculation for 8--12 singlet and triplet ESs using \textsc{gaussian16}, \cite{Gaussian16} so as to identify a series of relevant low-lying ESs. We next perform 
CC calculations for the targeted ESs with three basis sets, namely, {\Pop}, {\AVDZ}, and {\AVTZ}, which are dubbed as ``Pop'', ``AVDZ'', and ``AVTZ'' in the following. We underline that the selection of these three bases is motivated 
by their incorporation of both polarization and diffuse functions, in addition to their widespread availability. In particular, AVTZ is commonly acknowledged to yield valence excitation energies close to the complete basis set limit and 
provides reasonable VTE estimates for most Rydberg's ESs. Besides \textsc{gaussian}, CCSD and CC3 calculations are performed with \textsc{cfour} and \textsc{dalton}, \cite{dalton} the latter advantageously allowing CC3 triplet ES 
calculations. All CCSDT results presented here have been obtained with \textsc{cfour}. In \textsc{cfour}, converging the self-consistent field (SCF) procedure typically required a quadratic SCF algorithm together with fixed occupation numbers for each symmetry. 
In a few cases, we had to ramp up to $10^{-6}$ the linear dependency threshold to converge the SCF and CC calculations. Likewise, the default number of expansion vectors in the iterative CC subspace had to be increased to reach 
convergence of the CC ground-state amplitudes in some cases. For several ESs, we compared the CCSD/{\AVTZ} VTEs obtained with the three codes and did not observe any difference larger than 0.010 eV (a discrepancy obtained for 
a Rydberg  transition in anthracene) with the vast majority of VTEs agreeing to 0.001 eV.

To produce the theoretical best estimate (TBE) associated with each VTE, $\Delta {E}_{\text{AVTZ}}^{\text{TBE}}$, we typically select the CC3/AVTZ value when available, that is,
\begin{equation}
	\label{eq1}
	\Delta {E}_{\text{AVTZ}}^{\text{TBE}}  = 
	\Delta {E}_{\text{AVTZ}}^{\text{CC3}}.
\end{equation}
When we could afford to compute the CCSDT/Pop value, $\Delta E_{\text{Pop}}^{\text{CCSDT}}$, we relied on the following expression instead:
\begin{equation}
	\label{eq2}
	\Delta {E}_{\text{AVTZ}}^{\text{TBE}}  = \Delta E_{\text{AVTZ}}^{\text{CC3}} + \qty[ \Delta E_{\text{Pop}}^{\text{CCSDT}} - \Delta E_{\text{Pop}}^{\text{CC3}} ], 
\end{equation}
which has been shown to provide a reasonable estimate of the true $\Delta E_{\text{AVTZ}}^{\text{CCSDT}}$ value. \cite{Loo22}
In contrast, when CC3/AVTZ is beyond reach, which is typically the case for triplet ESs, we relied on the two following equations
\begin{eqnarray}
	\label{eq3}
	\Delta {E}_{\text{AVTZ}}^{\text{TBE}}  &=& 
	  \Delta E_{\text{AVTZ}}^{\text{CCSDR(3)}} + \qty[ \Delta E_{\text{AVDZ}}^{\text{CC3}} - \Delta E_{\text{AVDZ}}^{\text{CCSDR(3)}} ], \\
	\label{eq4}
	\Delta {E}_{\text{AVTZ}}^{\text{TBE}}  &=& 
	  \Delta E_{\text{AVTZ}}^{\text{CCSD}} + \qty[ \Delta E_{\text{AVDZ}}^{\text{CC3}} - \Delta E_{\text{AVDZ}}^{\text{CCSD}} ], 
\end{eqnarray}
to approach the true $\Delta {E}_{\text{AVTZ}}^{\text{CC3}}$ value, for singlets and triplets respectively. This composite procedure is quite popular in the CC community.\cite{Kal04,Bal06,Kam06b,Sil10b,Pea12,Wat12,Fel14,Gui18b,Fra19,Cas19,Chr21,Loo21a}
Indeed, it is common to correct double-$\zeta$ values obtained with a high excitation degree with triple-$\zeta$ estimates determined at a lower excitation level.
We refer the interested reader to Ref.~\citenum{Loo22} for exhaustive tests of this approach in a similar context.

In Tables S1-S4 of the SI, we provide, for all considered ESs, the raw VTEs computed at the CCSD, CCSDR(3), CC3, and CCSDT levels of theory with the three above-mentioned basis sets, together with several additional characteristics: 
i) the percentage of single excitations, $\%T_1$, as given at the CC3/AVDZ level by \textsc{dalton}; ii) the LR-CCSD/AVTZ oscillator strengths, $f$, provided by \textsc{gaussian}; iii) the spatial extend of the ES (and ground state), $\expval{r^2}$, determined 
at the ADC(2)/AVTZ level with \textsc{q-chem} 6.0 (see below for details); \cite{Epi21} and iv) the dominant Hartree-Fock (HF) molecular orbital contributions for each CCSD/AVTZ transition as computed with \textsc{gaussian}.

\subsection{Benchmarks}

The TBEs described above have been used to assess a series of popular wave function, density, and one-body Green's function methods. All these benchmark calculations were performed with the AVTZ basis set. 
Using the criteria detailed above, identifying the various ES across the benchmarked methods was generally tedious yet straightforward. All results are listed in the SI for each method detailed below.

\subsubsection{Wave Function Calculations}

We have used \textsc{turbomole} (7.3/7.5/7.8) \cite{Turbomole,Bal20} to determine the VTEs for 5 second-order methods, namely,  CIS(D), \cite{Hea94,Hea95} CC2,  \cite{Chr95,Hat00} SOS-ADC(2), SOS-CC2, 
and SCS-CC2.  \cite{Hel08} In these calculations, the resolution-of-the-identity (RI) approximation was systematically enforced, and default SOS and SCS parameters were applied. We have chosen 
\textsc{q-chem} \cite{Epi21} to perform the EOM-MP2, \cite{Sta95c} ADC(2), \cite{Tro97,Dre15} SOS-ADC(2), \cite{Kra13} and, when computationally feasible, ADC(3) \cite{Tro02,Har14,Dre15} calculations. These methods also take 
advantage of the RI implementation whereas the electron repulsion integral accuracy threshold was tightened to $10^{-14}$. \textsc{q-chem} and \textsc{turbomole} use different SOS parameters. Therefore, two non-equivalent SOS-ADC(2) methods 
have been benchmarked. These are distinguished by the additional labels, [QC] and [TM], for \textsc{q-chem} and \textsc{turbomole}, respectively. Within the composite ADC(2.5) approach proposed by some of us, \cite{Loo20b} VTEs were simply determined as the average of the ADC(2) and ADC(3) results. (Note that this simple half-and-half ratio was later shown to be nearly optimal by Dreuw's group. \cite{Bau22}) As explained above, \textsc{Gaussian16} \cite{Gaussian16} was 
used for the CCSD calculations. \cite{Pur82}  \textsc{dalton} \cite{dalton} was selected to computed the CCSDR(3) transition energies, \cite{Chr96b} whereas \textsc{cfour} was employed to compute the CCSD(T)(a)$^\star$ 
\cite{Mat16} and CCSDT-3 \cite{Wat96,Pro10} energies.

\subsubsection{TD-DFT Calculations}

\textsc{Gaussian16} was employed to perform the TD-DFT calculations using GHs and RSHs, except for a few cases where we relied on \textsc{turbomole} ([TM]) or \textsc{q-chem} ([QC]) instead (see below). 
We did not select any ``pure'' functionals, such as BLYP or PBE, as they are well-known to provide, compared to hybrids, less accurate VTEs for organic compounds. \cite{Lau13} The following 11 GHs were considered: 
TPSSh,  \cite{Sta03}  $\tau$-HCTCH-hyb, \cite{Boe02} B3LYP, \cite{Bec93,Fri94,Bar94,Ste94} PBE0, \cite{Ada99,Erz99} SCAN0 [TM], \cite{Hui16} M06,  \cite{Zha08b} SOGGA11-X,  \cite{Pev11b} BMK, \cite{Boe04}  
MN15,  \cite{Yu16a}  M08-HX,  \cite{Zha08c} and M06-2X. \cite{Zha08b} We also considered 11 RSHs, namely, M06-SX [TM], \cite{Wan20} CAM-B3LYP, \cite{Yan04} tCAM-B3LYP [TM], \cite{Oku12} mCAM-B3LYP [TM], 
\cite{Day06} rCAM-B3LYP [TM], \cite{Coh07} $\omega$B97X-D, \cite{Cha08b}  $\omega$B97M-V [QC], \cite{Mar16e} $\omega$B97X, \cite{Cha08} $\omega$B97, \cite{Cha08} LC-$\omega$HPBE, \cite{Hen09c} 
and M11. \cite{Pev11c} \textsc{Gaussian16} calculations use default parameters except for a tighter SCF convergence threshold ($10^{-10}$), \textsc{turbomole} calculations took advantage of the RI (and when possible RI-JK) approaches and 
we set a $10^{-9}$ SCF convergence threshold and a large quadrature grid (\texttt{gridsize=7}), whereas \textsc{q-chem} calculations were done without RI approximation and with tighten convergence parameters. \cite{xxx-QC}  
 
It is important to note that VTEs obtained with meta-GGA-based hybrids are known to be gauge-dependent. \cite{Bat12,Gro22} 
This effect is not accounted for in \textsc{gaussian}, although \textsc{q-chem} and 
\textsc{turbomole} propose a gauge invariance correction, the latter code computing it by default. Previous works have assessed the importance of this correction for small molecules, \cite{Gro22,Lia22,Gro23} 
as well as for ES properties of larger compounds. \cite{Gro23}  To evaluate the impact of the gauge invariance on VTEs, we have performed additional M06, M06-2X, and M06-SX calculations with \textsc{turbomole} (in addition to \textsc{gaussian} for the two former). To distinguish between these two sets of results, an additional prefix ``c'' is added to the name of the XCF. For example, cM06-2X corresponds to the gauge-corrected version of M06-2X.
  
\textsc{orca} 5.0  \cite{Nee20} was selected to perform the DH calculations using the \texttt{tightSCF} and \texttt{grid3} 
options and applying the RI with the automatically generated auxiliary basis set. The following 9 DHs were tested: B2PLYP, \cite{Gri06} PBE0-DH, \cite{Bre11} PBE-QIDH, \cite{Bre14} 
$\omega$B2PLYP, \cite{Cas19} RSX-0DH, \cite{Bre19} RSX-QIDH, \cite{Bre18b} $\omega$B97X-2, \cite{Cha09} SOS-$\omega$B88PP86, \cite{Cas21b} and SOS-$\omega$PBEPP86. \cite{Cas21b}

Finally, \textsc{turbomole} 7.8 was used for the local hybrid (LH) calculations. We used the same parameters as given above, including the large grid (\texttt{gridsize=7}) and considered 3 functionals, namely, LH12ct-SsirPW92,
\cite{Arb12} LH14t-calPBE, \cite{Arb14} and LH20t. \cite{Haa20} The above-mentioned gauge correction was applied. Hence, all these XCFs are denoted with the ``c'' prefix in the following.
 
For all XCFs, we performed two sets of calculations, one using the  ``full'' TD-DFT approach and one using the Tamm-Dancoff approximation (TDA). The only exception was the triplet ESs obtained with DHs that were 
determined with the TDA only. In case of triplet instabilities, \cite{Sea11,Pea11,Pea13} some codes, e.g., \textsc{turbomole}, simply stop the calculations whereas others go on (such as \textsc{gaussian}) and eventually print negative VTEs,
while some try to alleviate the problem by removing the vectors associated with the problematic states as, for example, \textsc{orca}. 
Because of these distinct behaviors, we have thus chosen to remove all occurrences of triplet instabilities.

\subsubsection{BSE/$GW$}

Full and TDA-based BSE calculations have been performed starting from both non-self-consistent $G_{0}W_{0}$ and eigenvalue-self-consistent ev$GW$ procedures. Comparisons of ev$GW$ with 
various alternative $GW$ procedures can be found elsewhere. \cite{Ran16c,Kap16} At the $GW$ level, we corrected all  occupied
(i.e., frozen core approximation is not applied here) and 50 virtual orbitals for each molecule, higher levels being rigidly shifted following the highest explicitly corrected level. These calculations have been 
conducted using \textsc{beDeft} (beyond DFT). \cite{Duc20,Duc21} The Coulomb-fitting RI (RI-V) techniques \cite{Ren12} and a robust analytic continuation scheme \cite{Duc21} have been employed during these 
calculations. The initial Kohn-Sham (KS) DFT (PBE0 and CAM-B3LYP) and HF calculations generating the starting eigenstates were performed with \textsc{orca}.\cite{Nee20}

\section{Results and Discussion}
\label{sec:Res}

\subsection{Theoretical Best Estimates}

Our TBEs are listed in Table \ref{Table-1}, representing a set of 122 VTEs, which can be categorized into various groups: i) 69 singlets and 53 triplets; ii) 83 $\ppi$ and 31 $\npi$ valence transitions as well as 8 Rydberg states; 
and  iii) 4 VTEs smaller than 2 eV, 13 in the 2--3 eV range, 46 in the 3--4 eV range, 39 in the 4--5 eV range, and 20 larger than 5 eV. These subgroups are quite representative of real-life chromophores typically investigated 
with TD-DFT, ADC(2), or CC2. The underrepresentation of Rydberg ESs in this set, compared to the QUEST database, \cite{Ver21} is a logical consequence of selecting large $\pi$-conjugated molecules, in which the majority 
of the lowest-lying ESs correspond to valence excitations.

\begin{table*}[htp]
\caption{\footnotesize Theoretical best estimates of the VTEs (in eV) for all considered ESs and the corresponding equation selected to define each of them.} 
\label{Table-1}
\vspace{-0.3 cm}
\footnotesize
\begin{tabular}{p{2.5cm}p{2.0cm}p{1.5cm}p{1.5cm}|p{2.5cm}p{2.0cm}p{1.5cm}p{1.5cm}}
ine
Compound	& State				&TBE	&Method		&Compound		& State				&TBE	&Method	\\
ine	
Anthracene	&$^1B_{1u}$ ($\ppi$)	&{3.757}&Eq.~\ref{eq1}	&				&$^3A_1''$ (Ryd)		&3.169	&Eq.~\ref{eq4}\\
			&$^1B_{2u}$ ($\ppi$)	&{3.782	}&Eq.~\ref{eq1}	&				&$^3E''$ (Ryd)			&3.693	&Eq.~\ref{eq4}\\
			&$^1B_{3g}$ ($\ppi$)	&{5.012	}&Eq.~\ref{eq1}	&Heptazine		&$^1A_2'$ ($\ppi$)		&2.717	&Eq.~\ref{eq2}\\
			&$^1B_{2g}$ (Ryd)		&5.078	&Eq.~\ref{eq3}	&				&$^1A_1''$ ($\npi$)		&3.999	&Eq.~\ref{eq2}\\
			&$^1B_{1u}$ ($\ppi$)	&{5.284}&Eq.~\ref{eq1}	&				&$^1E''$ ($\npi$)		&4.108	&Eq.~\ref{eq2}\\
			&$^1B_{3g}$ ($\ppi$)	&{5.291}&Eq.~\ref{eq3}	&				&$^1E'$ ($\ppi$)		&4.478	&Eq.~\ref{eq2}\\
			&$^1A_{g}$ ($\ppi$)		&{5.319}&Eq.~\ref{eq1}	&				&$^3A_2'$ ($\ppi$)		&2.936	&Eq.~\ref{eq3}\\
			&$^1B_{3u}$ (Ryd)		&5.379	&Eq.~\ref{eq3}	&				&$^3E''$($\ppi$)		&3.649	&Eq.~\ref{eq3}\\
			&$^1A_{u}$ (Ryd)		&5.451	&Eq.~\ref{eq3}	&				&$^3A_1''$ ($\npi$)		&3.992	&Eq.~\ref{eq3}\\
			&$^1B_{2u}$ ($\ppi$)	&{5.476}&Eq.~\ref{eq1}	&				&$^3E''$ ($\npi$)		&4.080	&Eq.~\ref{eq3}\\
			&$^3B_{1u}$ ($\ppi$)	&2.287	&Eq.~\ref{eq4}	&Naphthalimide	&$^1B_2$ ($\ppi$)		&4.033	&Eq.~\ref{eq1}\\
			&$^3B_{3g}$ ($\ppi$)	&3.655	&Eq.~\ref{eq4}	&				&$^1B_1$ ($\npi$)		&4.145	&Eq.~\ref{eq1}\\
			&$^3B_{2u}$ ($\ppi$)	&3.708	&Eq.~\ref{eq4}	&				&$^1A_1$ ($\ppi$)		&4.155	&Eq.~\ref{eq1}\\
Anthraquinone	&$^1B_{1g}$ ($\npi$)	&3.226	&Eq.~\ref{eq3}	&				&$^1A_2$ ($\npi$)		&4.632	&Eq.~\ref{eq1}\\
			&$^1A_{u}$  ($\npi$)		&3.466	&Eq.~\ref{eq3}	&				&$^3A_1$ ($\ppi$)		&2.780	&Eq.~\ref{eq4}\\
			&$^1A_{g}$  ($\ppi$)		&4.104	&Eq.~\ref{eq3}	&				&$^3B_2$ ($\ppi$)		&3.785	&Eq.~\ref{eq4}\\
			&$^1B_{2u}$ ($\ppi$)	&4.219	&Eq.~\ref{eq3}	&				&$^3B_1$ ($\npi$)		&4.060	&Eq.~\ref{eq4}\\
			&$^1B_{3g}$ ($\ppi$)	&4.321	&Eq.~\ref{eq3}	&				&$^3B_2$ ($\ppi$)		&4.124	&Eq.~\ref{eq4}\\
			&$^1B_{1u}$ ($\ppi$)	&5.101	&Eq.~\ref{eq3}	&				&$^3A_2$ ($\npi$)		&4.519	&Eq.~\ref{eq4}\\
			&$^1B_{2u}$ ($\ppi$)	&5.406	&Eq.~\ref{eq3}	&Naphthoquinone	&$^1B_1$ ($\npi$)		&3.023	&Eq.~\ref{eq2}\\
			&$^3B_{1g}$ ($\npi$)	&3.010	&Eq.~\ref{eq4}	&				&$^1A_2$ ($\npi$)		&3.215	&Eq.~\ref{eq2}\\
			&$^3A_u$ ($\npi$)		&3.262	&Eq.~\ref{eq4}	&				&$^1A_1$ ($\ppi$)		&4.115	&Eq.~\ref{eq2}\\
			&$^3B_{1u}$ ($\ppi$)	&3.443	&Eq.~\ref{eq4}	&				&$^1B_2$ ($\ppi$)		&4.342	&Eq.~\ref{eq2}\\
			&$^3B_{3g}$ ($\ppi$)	&3.498	&Eq.~\ref{eq4}	&				&$^1A_2$ ($\npi$)		&5.436	&Eq.~\ref{eq2}\\	
			&$^3A_g$ ($\ppi$)		&3.740	&Eq.~\ref{eq4}	&				&$^1A_1$ ($\ppi$)		&5.458	&Eq.~\ref{eq2}\\	
Aza-BODIPY	&$^1B_2$ ($\ppi$)		&2.512	&Eq.~\ref{eq2}	&				&$^1B_2$ ($\ppi$)		&5.548	&Eq.~\ref{eq2}\\	
			&$^1B_2$ ($\ppi$)		&3.457	&Eq.~\ref{eq2}	&				&$^3B_1$ ($\npi$)		&2.796	&Eq.~\ref{eq4}\\	
			&$^1A_1$ ($\ppi$)		&3.463	&Eq.~\ref{eq2}	&				&$^3A_2$ ($\npi$)		&3.009	&Eq.~\ref{eq4}\\	
			&$^1B_1$ ($\npi$)		&3.881	&Eq.~\ref{eq2}	&				&$^3B_2$ ($\ppi$)		&3.316	&Eq.~\ref{eq4}\\
			&$^3B_2$ ($\ppi$)		&1.329	&Eq.~\ref{eq4}	&				&$^3B_2$ ($\ppi$)		&3.445	&Eq.~\ref{eq4}\\
			&$^3B_2$ ($\ppi$)		&2.774	&Eq.~\ref{eq4}	&				&$^3A_1$ ($\ppi$)		&3.760	&Eq.~\ref{eq4}\\
			&$^3A_1$ ($\ppi$)		&3.027	&Eq.~\ref{eq4}	&Phenazine		&$^1B_{1u}$ ($\npi$)	&3.374	&Eq.~\ref{eq1}\\
			&$^3B_1$ ($\npi$)		&3.435	&Eq.~\ref{eq4}	&				&$^1B_{3u}$ ($\ppi$)	&3.717	&Eq.~\ref{eq1}\\
Azobenzene	&$^1B_g$ ($\npi$)		&2.871 	&Eq.~\ref{eq1}	&				&$^1B_{2u}$ ($\ppi$)	&3.744	&Eq.~\ref{eq1}\\
			&$^1B_u$ ($\ppi$)		&4.231	&Eq.~\ref{eq1}	&				&$^1B_{1g}$ ($\ppi$)	&4.497	&Eq.~\ref{eq1}\\
			&$^1A_g$ ($\ppi$)		&4.439	&Eq.~\ref{eq1}	&				&$^1B_{2g}$ ($\npi$)	&4.837	&Eq.~\ref{eq1}\\
			&$^1B_u$ ($\ppi$)		&4.446	&Eq.~\ref{eq1}	&				&$^1A_u$ ($\ppi$)		&5.115	&Eq.~\ref{eq1}\\	
			&$^1A_g$ ($\ppi$)		&5.185	&Eq.~\ref{eq1}	&				&$^3B_{3u}$ ($\ppi$)	&2.408	&Eq.~\ref{eq4}\\
			&$^3B_g$ ($\npi$)		&2.286	&Eq.~\ref{eq4}	&				&$^3B_{1u}$ ($\npi$)	&3.055	&Eq.~\ref{eq4}\\
			&$^3B_u$ ($\ppi$)		&2.800	&Eq.~\ref{eq4}	&				&$^3B_{2u}$ ($\ppi$)	&3.376	&Eq.~\ref{eq4}\\	
			&$^3A_g$ ($\ppi$)		&3.921	&Eq.~\ref{eq4}	&				&$^3B_{1g}$ ($\ppi$)	&3.554	&Eq.~\ref{eq4}\\	
			&$^3B_u$ ($\ppi$)		&4.218	&Eq.~\ref{eq4}	&Phthalimide		&$^1B_1$ ($\npi$)		&4.186	&Eq.~\ref{eq2}\\	
			&$^3A_g$ ($\ppi$)		&4.279	&Eq.~\ref{eq4}	&				&$^1A_1$ ($\ppi$)		&4.600	&Eq.~\ref{eq2}\\
BODIPY		&$^1B_2$ ($\ppi$)		&{2.771}&Eq.~\ref{eq2}	&				&$^1A_2$ ($\npi$)		&4.795	&Eq.~\ref{eq2}\\
			&$^1B_2$ ($\ppi$)		&{3.804	}&Eq.~\ref{eq2}	&				&$^1B_2$ ($\ppi$)		&4.942	&Eq.~\ref{eq2}\\
			&$^1A_1$ ($\ppi$)		&{3.901	}&Eq.~\ref{eq2}	&				&$^1B_2$ ($\ppi$)		&5.921	&Eq.~\ref{eq2}\\
			&$^3B_2$ ($\ppi$)		&1.862	&Eq.~\ref{eq4}	&				&$^1A_2$ ($\npi$)		&5.999	&Eq.~\ref{eq2}\\
			&$^3B_2$ ($\ppi$)		&3.110	&Eq.~\ref{eq4}	&				&$^1A_1$ ($\ppi$)		&6.306	&Eq.~\ref{eq2}\\
			&$^3A_1$ ($\ppi$)		&3.339	&Eq.~\ref{eq4}	&				&$^3B_2$  ($\ppi$)		&3.762	&Eq.~\ref{eq1}\\
Coumarin		&$^1A'$ ($\ppi$)		&4.307	&Eq.~\ref{eq1}	&				&$^3B_1$ ($\npi$)		&3.957	&Eq.~\ref{eq1}\\
			&$^1A''$ ($\npi)$		&4.796	&Eq.~\ref{eq1}	&				&$^3A_1$  ($\ppi$)		&4.344	&Eq.~\ref{eq1}\\
			&$^1A'$ ($\ppi$)		&4.980	&Eq.~\ref{eq1}	&				&$^3B_2$  ($\ppi$)		&4.506	&Eq.~\ref{eq4}\\
			&$^3A'$ ($\ppi$)		&3.264	&Eq.~\ref{eq4}	&				&$^3A_2$  ($\npi$)		&4.578	&Eq.~\ref{eq1}\\
			&$^3A'$ ($\ppi$)		&4.109	&Eq.~\ref{eq4}	&				&$^3B_2$  ($\ppi$)		&4.666	&Eq.~\ref{eq4}\\
			&$^3A''$ ($\npi)$		&4.655	&Eq.~\ref{eq4}	&Tolan			&$^1B_{2u}$ ($\ppi$)	&4.724	&Eq.~\ref{eq3}\\
Cyclazine		&$^1A_2'$ ($\ppi$)		&0.979	&Eq.~\ref{eq2}	&				&$^1B_{3g}$ ($\ppi$)	&4.746	&Eq.~\ref{eq3}\\
			&$^1E'$ ($\ppi$)		&3.018	&Eq.~\ref{eq2}	&				&$^1B_{1u}$ ($\ppi$)	&4.796	&Eq.~\ref{eq3}\\
			&$^1A_1''$ (Ryd)		&3.163	&Eq.~\ref{eq2}	&				&$^1A_u$ ($\ppi$)		&5.546	&Eq.~\ref{eq3}\\
			&$^1E''$ (Ryd)			&3.691	&Eq.~\ref{eq2}	&				&$^1B_{3u}$ (Ryd)		&5.598	&Eq.~\ref{eq3}\\
			&$^3A_2'$ ($\ppi$)		&1.110	&Eq.~\ref{eq4}	&				&$^3B_{1u}$ ($\ppi$)	&3.261	&Eq.~\ref{eq4}\\
			&$^3E'$ ($\ppi$)		&2.154	&Eq.~\ref{eq4}	&				&$^3A_g$ ($\ppi$)		&4.061	&Eq.~\ref{eq4}\\
ine
\end{tabular}
\end{table*}

\subsubsection{{On the accuracy of the TBEs}}

While CC3 demonstrated exceptional accuracy for the compounds included in the QUEST database, with a trifling MAE of 0.02 eV, \cite{Ver21} predicting changes in its accuracy for the larger compounds 
studied here is challenging. Thus, we do not provide an anticipated error bar for the TBEs listed in Table \ref{Table-1}, though we are confident that these values are the most trustworthy available 
for the systems depicted in Figure \ref{Fig-1}.

To delve deeper into the accuracy of the TBEs, a valuable metric is $\%T_1$, which defines the percentage of single-excitation character of the transition. Indeed, systematically increasing the degree of the CC 
excitation operator would not be the optimal strategy if doubly-excited  states (characterized by low values of $\%T_1$) were involved. \cite{Loo19c,Kos24} Among the 122 cases treated here, 62 show $\%T_1 > 90\%$, 
indicating that they can be safely viewed as almost pure single excitations. Notably, this includes all 53 triplets, which comes as no surprise. \cite{Sch08,Ver21} Most of the remaining singlet states, i.e., 44, fall in 
the $90\% > \%T_1 > 85\%$  range. {In other words, 106-out-of-122 TBEs are associated with $\%T_1 > 85\%$.  For these transitions, both CC3 and CCSDT generally performs extremely well.} Indeed, 
 for such transitions we demonstrated, {on the one hand},  that corrections from quadruples (CC4 or CCSDTQ) are essentially negligible, \cite{Loo22} {whereas, on the other hand, second-order 
multireference methods, e.g. CASPT2 and NEVPT2, are significantly less accurate than CC3 with errors of approximatively 0.10-0.13 and 0.02 eV, respectively.}\cite{Sar22} Next, 10 states are characterized 
by $85\% > \%T_1 > 80\%$, for which both CC3 and CCSDT are likely no longer chemically accurate, but probably still superior to CASPT2 (or similar methods), based on our experience. \cite{Sar22,Bog22}
Finally, Table \ref{Table-1} includes a few transitions with $80\% > \%T_1 > 75\%$ (6 ESs). In these cases, we expect the present TBEs to be too large, i.e., upper bounds of the exact VTEs, as corrections from 
quadruples typically downshift such states with a significant double excitation character. \cite{Loo22} {For such transitions in rather large compounds, it is difficult to know if the CC3/CCSDT or CASPT2/NEVPT2 
would be the most accurate. We can only refer to smaller structures for which indisputable TBEs are at hand.} \cite{Ver21,Kos24} {One can obtain hints from smaller compounds.} For the well-known $^1A_g$ dark 
ES of butadiene ($\%T_1 = 75\%$), the CC3 and CCSDT errors, relative to the near-exact VTE, are +0.15 and +0.08 eV, respectively, {whereas the CASPT2 and NEVPT2 deviations are +0.22 and +0.18 eV,
respectively.} \cite{Ver21,Kos24} {For the lowest $^1E_{2g}$ ES of benzene ($\%T_1 = 73\%$), the CC3, CASPT2 and NEVPT2 errors are respectively +0.18, +0.12, and +0.32 eV.}\cite{Kos24} {Consequently,
there is no clear indication that multireference values would be more accurate than the present CC-based TBEs for our set of data.}

\begin{figure}[htp]
  \includegraphics[width=\linewidth]{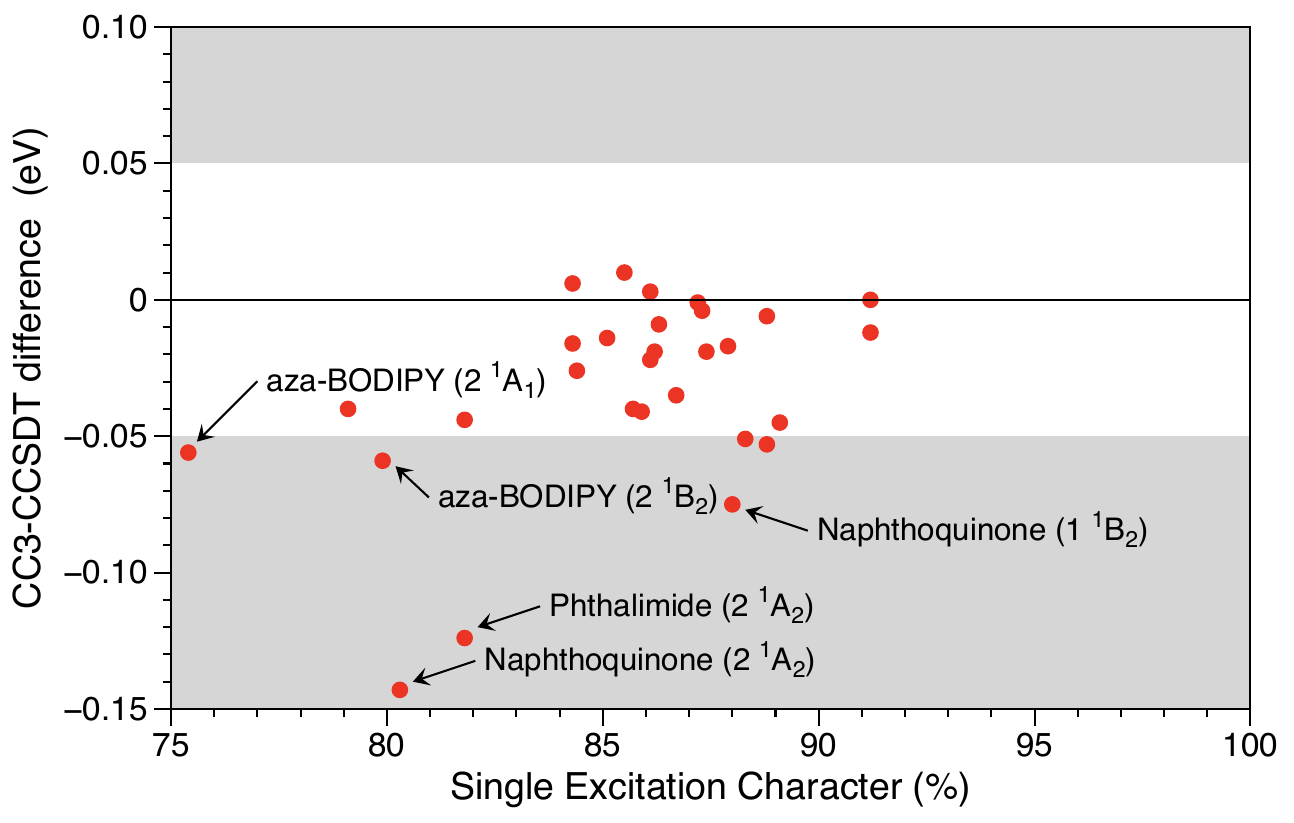}
  \caption{Relationship between the CC3-CCSDT VTE differences (in eV) and the single-excitation character (computed at the CC3 level). The central white zone indicate absolute errors smaller than 0.05 eV. 
  Some outliers are explicitly indicated.
}
  \label{Fig-2}
\end{figure}

Another strategy to probe CC3's accuracy is to investigate the differences between the CC3 and CCSDT values. Though this approach is not entirely foolproof {in assessing the impact of further (quadruple)
CC corrections}, \cite{Loo22} it hints at the convergence of the CC family of methods. From the {29} singlet ESs for which we could compute this difference in the {\Pop} basis set, it results a MAE of 0.034 eV
which we regard as a positive indicator of our TBEs' quality. {We note that the use of this small basis set should not be an issue. Indeed, we have shown that Eq.}~\ref{eq2} {provides excellent estimates of the
actual CCSDT/AVTZ results, with a MAE of 0.005 eV, i.e., the CC3-CCSDT differences seems reasonably insensitive to the basis set size.}\cite{Loo22} Figure \ref{Fig-2} shows {these differences} as a function 
of $\%T_1$. As can be seen, there is no clean relationship between the two parameters, although the two largest deviations correspond to states with  $\%T_1 < 85\%$.  Another interesting outcome, already 
noticed for substituted phenyls, \cite{Loo24} is that the CCSDT correction tends to be positive, meaning that the CC3 VTEs tend to be lower than their  CCSDT counterparts for large molecules. 

\subsubsection{{Comparisons with literature data}}

Let us now briefly compare the TBEs gathered in Table \ref{Table-1} to previous literature values. While a plethora of TD-DFT data, as well as numerous ADC(2) and CC2 results, are available for many chromophores, 
we decided to focus on data obtained with higher levels of theory. Likewise, we eschewed including experimental data since the present VTEs do not account for vibronic or solvent effects, 
making well-grounded comparisons with measured UV/Vis spectra of solvated dyes unfeasible. For most molecules, we could not find any previous highly-accurate estimates of these VTEs.

For anthracene, one can find investigations performed with CASPT2 and DMRG, \cite{Kur14,Bet17,Sch18c,Sha19c} most of them focussing on the singlet-triplet or $L_a$-$L_b$ gaps. 
Reference \citenum{Kur14} reports the following DMRG-CASPT2/cc-pVTZ(-f) transition energies: 1.73 eV ($^3B_{1u}$), 3.45 eV ($^1B_{1u}$ ), 3.77 eV ($^1B_{2u}$), 4.97 eV ($^1A_{g}$), and 
5.00 eV ($^1B_{3g}$).  There is globally a fair agreement with the current TBEs that are typically larger, especially for the lowest triplet ES, for which the agreement is especially poor.
Note that for the $^1A_g$ ES of anthracene, we have $\%T_1 = 75\%$. Therefore, it is likely that the VTE reported in Table \ref{Table-1} is an upper bound of the exact value.


For aza-BODIPY, the 2.34 eV value obtained at the CASPT2/cc-pVDZ level by Momeni and Brown for the lowest singlet transition \cite{Mom15} is slightly below the present TBE of 2.51 eV.

For azobenzene, one can find  MS-CASPT2/6-31G(d) \cite{Cas16}, SS-RASPT2/VDZP, \cite{Ale19} and CC3/{\AVDZ} \cite{Hut20} results for the two lowest ESs. These methods respectively return 
VTEs of 2.86, 2.66, and 2.91 eV for the $\npi$ transitions and 4.17, 3.86, and 4.07 eV for the $\ppi$ transitions. These previous data are reasonably close to the present estimates of 2.87 and 4.23 eV. 
We note that the MLCC3/{\AVDZ} results of Ref.~\citenum{Pau22} have been determined on a significantly distorted geometry precluding sound comparisons with the present work.

A very complete study of the photophysics of BODIPY was performed by De Vetta and coworkers. \cite{Dev19} For the vertical absorption, they reported CASPT2/ANO-L transition energies of
2.64, 3.78, and 3.87 eV for the three lowest singlet ESs, and 1.92, 3.11, and 3.31 eV for the three lowest triplet ESs. These values are globally in excellent agreement with the TBEs of Table \ref{Table-1}
with a MAE of 0.04 eV only. One can also find earlier CASPT2 estimates with smaller basis sets for the lowest singlet transition of BODIPY. \cite{Bri13,Mom15}

The CASPT2/6-31G(d) transition energies to the three lowest singlet ESs are 4.41, 4.66, and 5.13 eV in coumarin. \cite{Gan17} For the first and third states, of $\ppi$ nature, these values are slightly
larger than the present estimates of 4.31 and 4.98 eV, whereas the shift is in the other direction for the second state, corresponding to a $\npi$ transition, for which the present TBE is 4.80 eV. 

For the lowest singlet and triplet ESs of both cyclazine and heptazine, which notably exhibit a negative singlet-triplet gap, previous SC-NEVPT2/def2-TZVP estimates reported VTEs of
1.22 eV (singlet) and 1.28 eV (triplet) for cyclazine and 3.26 eV (singlet) and 3.40 eV (triplet) for heptazine. \cite{Ric21} These values are significantly upshifted compared to the present results.
There are other multi-reference values available, \cite{Gho22,Tuc22,Drw23} and one can find a complete list of literature values  for the lowest-lying transitions in Tables 1 and 2 of Ref.~\citenum{Loo23a}.

	



For tolan, the study of Robertson and Worth provides CASPT2/cc-pVDZ VTEs of 4.71 eV ($^1B_{2u}$), 4.72 eV ($^1B_{3g}$), 5.04 eV ($^1B_{1u}$), and 6.05 eV ($^1A_u$). \cite{Rob18b} The two
former are very close to the current TBEs.

\subsection{Wave Function Benchmarks}

\begin{figure*}[htp]
  \includegraphics[scale=.9,viewport=2cm 14.5cm 19cm 27.5cm,clip]{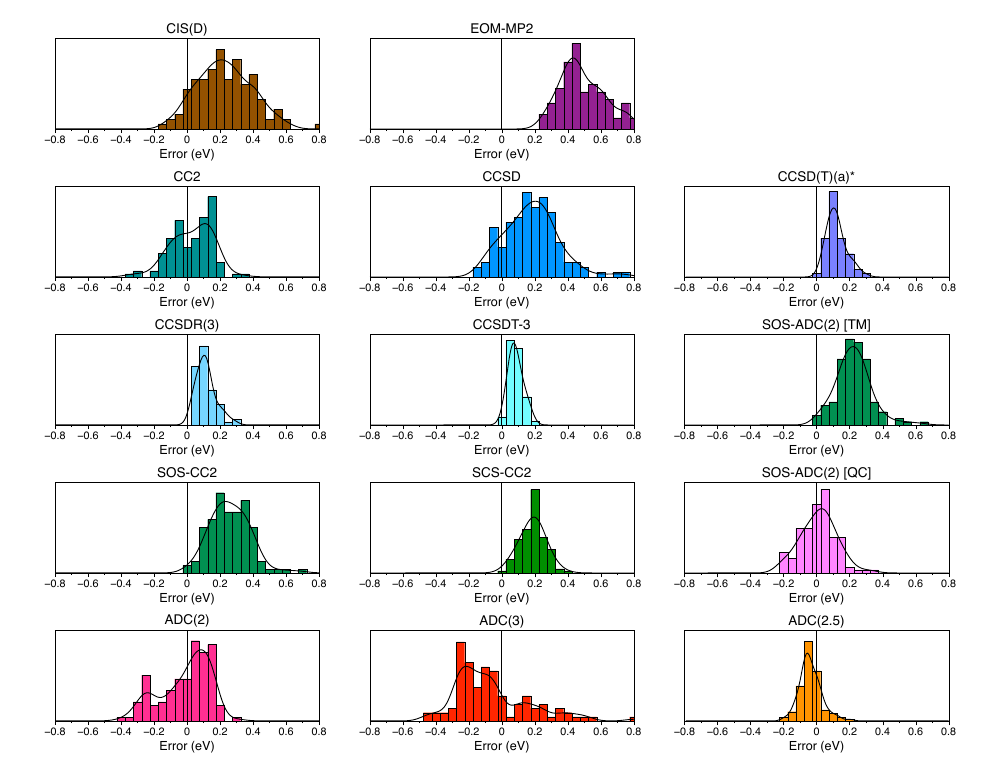}
  \caption{Histograms of the errors as compared to the TBEs for wave function methods considered in the present work.
  }
  \label{Fig-3}
\end{figure*}

Having the above-described TBEs at hand, it is natural to assess the performances of lower-level wave function approaches. Indeed, the CC3/{\AVTZ} VTEs listed in Table \ref{Table-1} are
at the frontier of today's technical feasibility. Thus, for practical applications, one would clearly wish to use methods that are computationally more affordable. The interested reader can find
the full list of raw data in the SI for the 14 wave function methods tested here.  In Table \ref{Table-2}, we report the results of the statistical analysis: mean signed error (MSE), MAE, and standard 
deviation of the errors (SDE), together with the MAE obtained for various subgroups of ESs. In {Figures} \ref{Fig-3}, {S1, and S2,} histograms representing the distribution of these errors 
are available.  It should be recalled that: i) there are only 8 Rydberg transitions in our set; ii) CCSD(T)(a)$^\star$, CCSDR(3), and CCSDT-3 have been implemented for singlet ESs only; and 
iii) for some of the most expensive levels of theory [CCSDT-3 for instance], we were not able to obtain all possible VTEs despite our efforts.


\begin{table}[htp]
\caption{\footnotesize Statistical analysis of the performance of various wave function methods. All values are in eV}
\label{Table-2}
\vspace{-0.3 cm}
\footnotesize
\begin{tabular}{lccccccccc}
ine
	    & \multicolumn{3}{c}{Full set}	& \multicolumn{6}{c}{MAE of subsets}\\
Method & MSE & MAE & SDE & \begin{sideways}Singlet\end{sideways} & \begin{sideways}Triplet\end{sideways} &  \begin{sideways}$\ppi$\end{sideways} & 
 \begin{sideways}$\npi$\end{sideways} &  \begin{sideways} Rydberg \end{sideways} &  \begin{sideways} $\%T_1 \geq 85\%$ \end{sideways}  \\
ine
CIS(D)				&0.24	&0.24	&0.18	&0.22	&0.28	&0.29	&0.18	&0.06	&0.23\\
CC2					&0.03	&0.11	&0.12	&0.09	&0.12	&0.11	&0.10	&0.07	&0.10\\
EOM-MP2				&0.51	&0.51	&0.16	&0.59	&0.41	&0.54	&0.47	&0.40	&0.47\\
CCSD				&0.18	&0.20	&0.16	&0.27	&0.10	&0.18	&0.27	&0.04	&0.16\\
CCSD(T)(a)$^\star$		&0.12	&0.12	&0.07	&0.12	&		&0.11	&0.15	&0.03	&0.10\\
CCSDR(3)			&0.11	&0.11	&0.07	&0.11	&		&0.11	&0.14	&0.03	&0.09\\
CCSDT-3				&0.08	&0.08	&0.05	&0.08	&		&0.08	&0.11	&0.01	&0.07\\
SOS-ADC(2)$^\text{[TM]}$&0.22	&0.22	&0.11	&0.21	&0.24	&0.20	&0.28	&0.24	&0.21\\
SOS-CC2				&0.26	&0.26	&0.12	&0.25	&0.28	&0.22	&0.38	&0.17	&0.24\\
SCS-CC2				&0.19	&0.19	&0.07	&0.17	&0.21	&0.18	&0.22	&0.09	&0.18\\
SOS-ADC(2)$^\text{[QC]}$&0.01	&0.09	&0.11	&0.10	&0.06	&0.08	&0.08	&0.12	&0.08\\
ADC(2)				&-0.01	&0.12	&0.15	&0.11	&0.14	&0.10	&0.21	&0.04	&0.12\\
ADC(3)				&-0.06	&0.20	&0.22	&0.18	&0.23	&0.18	&0.24	&0.19	&0.19\\
ADC(2.5)				&-0.04	&0.06	&0.06	&0.06	&0.06	&0.06	&0.05	&0.06	&0.06\\
ine
\end{tabular}
\end{table}

In the rightmost column of Table \ref{Table-2}, we provide the MAEs obtained for ESs with a single-excitation character, as defined by $\%T_1 \geq 85\%$. As discussed above, 
we anticipate a very small CC3 error ($\leq$ 0.05 eV) for the vast majority of these states. \cite{Loo22} As can be seen, the MAEs obtained for this subset are only slightly smaller than 
their counterparts determined on the full set: the typical variations are below 0.04 eV and the methodological trends are preserved. Therefore, we have chosen to perform the following analyses 
using the entire set of TBEs.

Both CIS(D) and EOM-MP2 provide quite poor transition energies, with a clear overshooting trend as well as large SDEs. This is no surprise. \cite{Goe10a,Jac15b,Loo18a,Loo20a,Ver21,Loo21,Loo24}
Practically, these approaches are therefore of limited interest: their performances are not clearly superior to the best GHs, while they appear significantly less accurate than several DHs presenting the same 
$\mathcal{O}(N^5)$ formal scaling with system size (\emph{vide infra}). The EOM-MP2 MAE of 0.51 eV is much larger than its QUEST counterpart of 0.22 eV, and even exceeds the MAE  obtained for 
molecules encompassing 7--10 non-H atoms belonging to the QUEST database (0.42 eV).  \cite{Ver21}  This indicates that the EOM-MP2 error continues to grow with system size. 
Nevertheless, in the present set, this overestimation remains relatively systematic, as evidenced by the SDE. In contrast, CIS(D) shows errors in the 0.2--0.3 eV range for all molecular 
sizes. In other words, it is less affected by the physical extent of the investigated compounds. These errors are notably smaller for both Rydberg and $\npi$ transitions than for $\ppi$ excitations, 
consistently with earlier findings. \cite{Ver21}

For the present set of real-life dyes, CC2 emerges as particularly robust with a small MSE, an attractive SDE of 0.12 eV (indicating quite systematic errors) and MAE close to 0.10 eV for all ES families. 
Interestingly, based on QUEST, we found that the MAE of CC2 is 0.16 eV, \cite{Ver21} but this larger value was mostly driven by the small compounds. Indeed, the MAE is 0.11 eV for the 7--10 (non-H) atom 
molecules of QUEST, hinting that the convergence of the CC2 error with system size is likely reached.  {Nevertheless, one notes a bimodal distribution of the CC2 errors in Figure} \ref{Fig-3}. {As
can be seen in Figure S1, this originates from the tendency of CC2  to underestimate (overestimate) the VTE of both Rydberg and $\npi$ ($\ppi$) ESs, which might cause issues if a balanced description of all
state types is required.} For comparisons, the MAE and SDE of CC2 were found to be 0.07 eV and 0.09 eV, respectively, in the study of Winter and coworkers, \cite{Win13} who selected
0-0 experimental energies (for the lowest-lying ESs) as references.  We believe that our statistical indicators are slightly higher due to the inclusion of more challenging (higher-energy) ESs in our set.

As discussed in many works, \cite{Hat05c,Win13,Har14,Jac15b,Kan17,Loo18a,Loo20a,Ver21,Loo21,Loo24} ADC(2) does provide a rather similar accuracy as CC2. However, ADC(2) seems slightly less 
effective than its CC counterpart here, as evidenced by the higher values of the statistical indicators for most ES subgroups considered in Table \ref{Table-2}, especially the $\npi$ transitions for which the 
advantage of CC2 over ADC(2) is clear, {since ADC(2) more significantly underestimates the VTEs of those transitions than CC2} ({See Figures S1 and S2}). This finding differs 
from the one of Ref.~\citenum{Ver21} but parallels the outcomes of Winter and coworkers, \cite{Win13}  who reported a MAE of 0.08 eV and a SDE of 0.12 eV for ADC(2).

Even though the overestimation tendency of CCSD is well-documented, \cite{Sch08,Car10,Wat13,Kan14,Kan17,Dut18,Ver21,Loo21,Loo24} the MSE (0.18 eV) and MAE (0.20 eV) of 
Table \ref{Table-2} are surprisingly large, with an obvious unbalanced nature. {In particular, the CCSD positive biais is markedly larger for singlets than for triplets (Figure S1)}. It is also larger 
for $\npi$ than for $\ppi$ transitions, {though in that case the error pattern seem better than their CC2 counterparts}. What was even less expected is the large SDE of 0.16 eV, which clearly exceeds
its CC2 counterpart of 0.12 eV.  Indeed, the reverse SDE trend was noticed in QUEST, \cite{Ver21} CCSD delivering more systematic errors than CC2. In light of these outcomes, the interest in using 
CCSD rather than CC2 for applications on large compounds is {difficult to justify in general}.  Adding perturbative corrections for the missing triple excitations helps, with SDE well below 0.10 eV 
for both CCSD(T)(a)$^\star$ and CCSDR(3). The MSEs of these two similar methods remain firmly positive, though. Unfortunately, this confirms \cite{Ver21,Loo24} that these two schemes become 
less effective when one considers larger molecules as they cannot downshift the CCSD VTEs enough. To solve this issue, one can turn to iterative triples with CCSDT-3, but this expensive method 
shares with CC3 the unpleasant $\mathcal{O}(N^7)$ scaling with system size. Note that CCSDT-3 still overestimates our TBEs, albeit by a small acceptable margin.

As originally reported by us, \cite{Loo20b} ADC(3) does not improve upon ADC(2), while ADC(2.5), where one simply average the ADC(2) and ADC(3) VTEs, \cite{Ver21,Bau22,Loo24} appears again to be a 
relatively affordable $\mathcal{O}(N^6)$ approach providing accurate results.  {Figure S2 clearly shows that the error patterns of ADC(2) and ADC(3) are mirrored for the different
types of ESs, explaning the success of ADC(2.5). Indeed, all} statistical errors of {ADC(2.5-} are close to 0.06 eV and it impressively outperforms both CCSD(T)(a)$^\star$ 
and CCSDR(3) on the present set, with an absolute accuracy comparable to the one of CCSDT-3, but with an underestimation trend.

Briefly, the most effective spin-scaled approach evaluated here is likely SOS-ADC(2) with the QC parameters, \cite{Kra13} which delivers an accuracy comparable to the one of CC2, for the same
formal scaling but with a smaller prefactor. SCS-CC2 (with the TM parameters) yields a very small SDE but at the cost of a clear overshooting trend, a conclusion paralleling previous studies. 
\cite{Goe10a,Jac15b,Taj20a,Loo20d,Ver21,Loo21,Loo24}  Both the SOS-ADC(2) and SOS-CC2 with TM parameters appear as less reliable options for the current set. {All tested spin-scaled
methods also provide reasonably balanced error patterns for singlets and triplets (Figure S2), which is likely advantageous when estimating singlet-triplet gaps.}

\subsection{TD-DFT and BSE/$GW$ Benchmarks}

\subsubsection{Gauge Invariance Correction}
\label{sec:gauge}

As mentioned in Section \ref{sec:Comp}, one can include gauge invariance corrections in TD-DFT calculations involving (hybrid) meta-GGA XCFs. The impact of these corrections was carefully studied and quantified
in previous works. \cite{Gro22,Lia22,Gro23} As enforcing the TDA breaks gauge invariance, it has been argued that applying the gauge corrections at the TDA-DFT level is not necessarily well-grounded. \cite{Lia22} 
Nevertheless, we have found that the magnitude of the corrections obtained with TDA-DFT and ``full'' TD-DFT for a given transition are extremely similar (see Figures {S3 and S4} in the SI).

\begin{figure}[htp]
  \includegraphics[width=\linewidth]{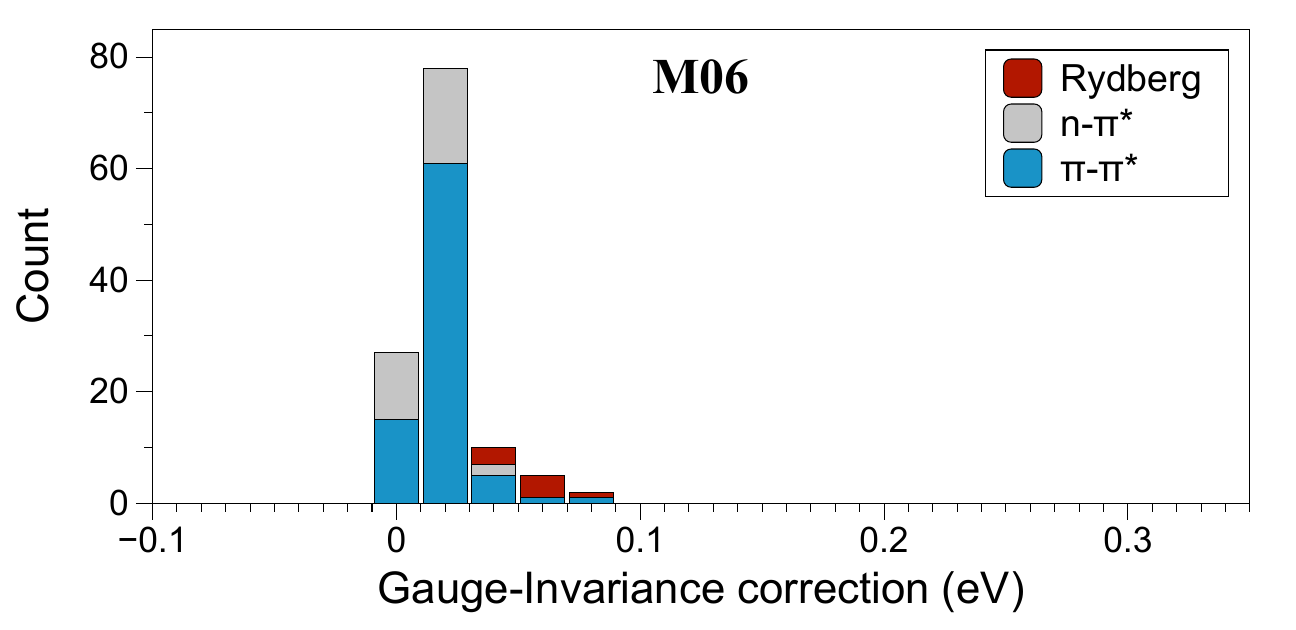}
  \includegraphics[width=\linewidth]{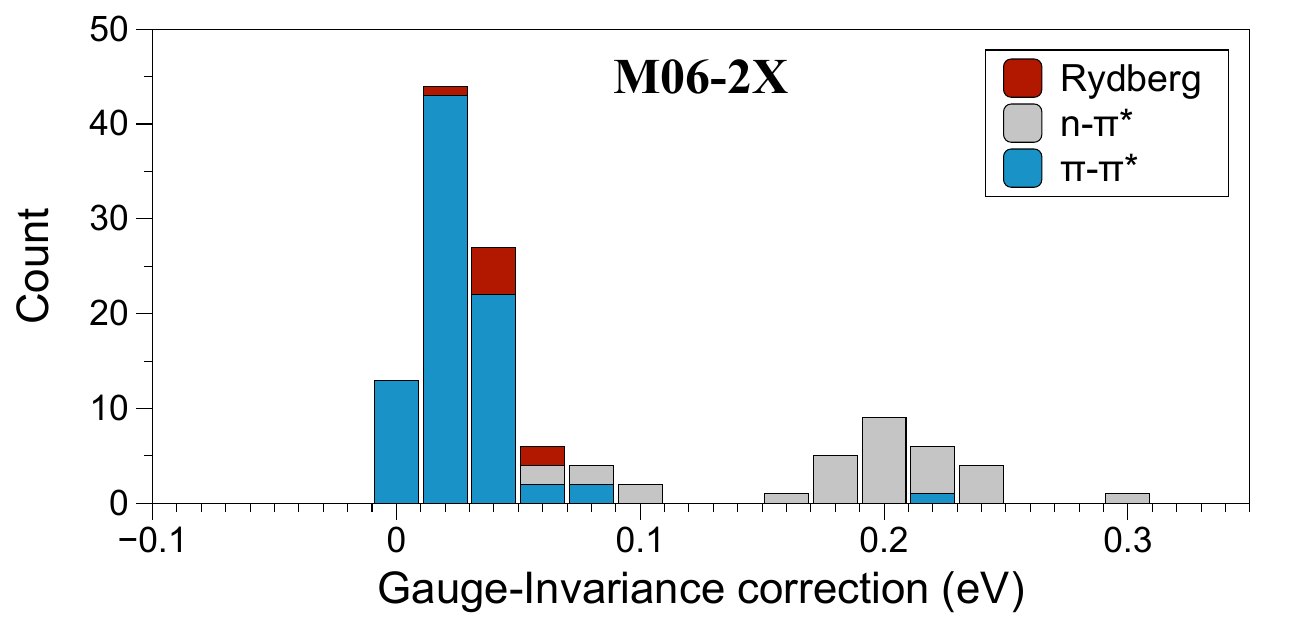}
  \includegraphics[width=\linewidth]{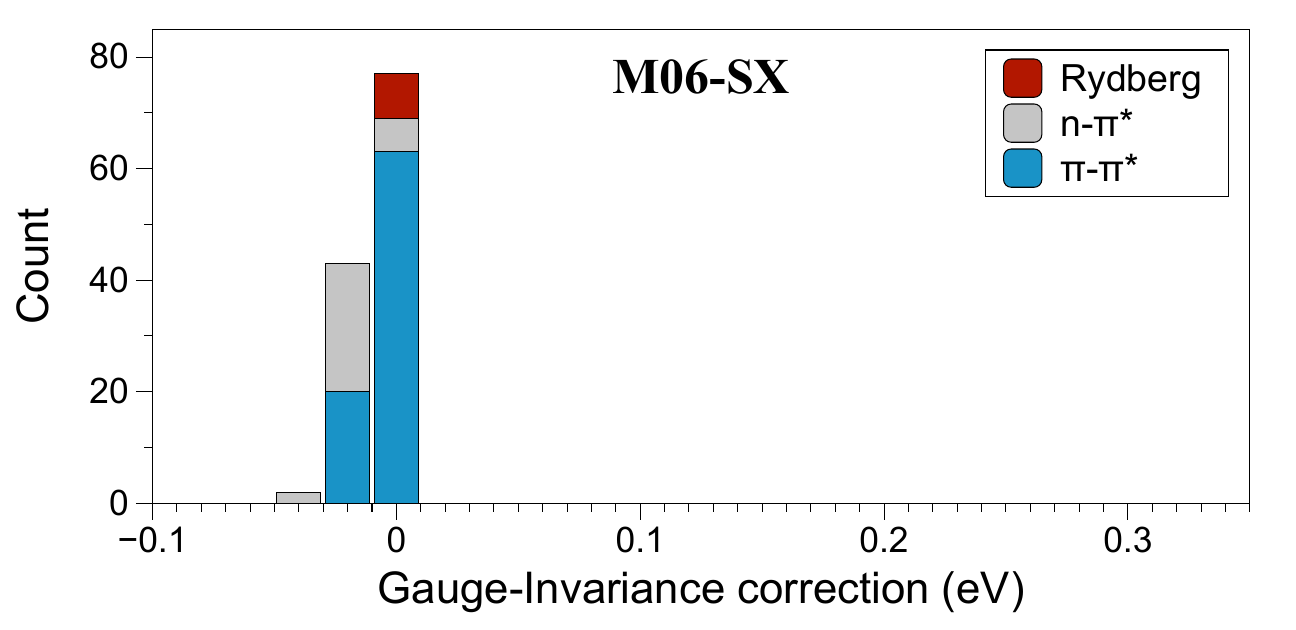}
  \caption{Impact of the gauge invariance corrections for various functionals at the TD-DFT level.}
  \label{Fig-4}
\end{figure}

Histograms of the changes in VTEs induced by the gauge invariance corrections can be found in Figure \ref{Fig-4} for M06, M06-2X, and M06-SX. We note that the corrections are positive with the first two 
XCFs (i.e., they tend to increase the VTEs) but are negative for M06-SX. This trend was expected for M06 and M06-2X, and a rationale for the sign of the correction can be found elsewhere. \cite{Gro22} 
The average changes are negligible with both M06 (+0.02 eV on average) and M06-SX ($-0.01$ eV on average). These values hold for both TDA-DFT and TD-DFT. With M06, the increase is slightly larger for the Rydberg ESs 
(+0.05 eV on average) than for the valence transitions (+0.02 eV for $\ppi$ and +0.01 eV for $\npi$), whereas variations are always minimal for M06-SX. The impact of the gauge invariance correction becomes more 
significant for M06-2X, with average variations of +0.03 eV ($\ppi$), +0.18 eV ($\npi$), and +0.04 eV (Rydberg). We could not detect any significant difference between singlet and triplet ESs regarding the magnitude
of these corrections. The M06-2X values are very similar to the one reported by Grotjahn \emph{et al.}~for smaller molecules but with the same XCFs:  +0.04 eV ($\ppi$) and +0.17 eV ($\npi$). \cite{Gro22}  There is however one clear 
outlier in our set: the $^1A_u$ $\ppi$ transition of tolan that undergoes a change of +0.17 eV (TDA-DFT) or +0.22 eV (TD-DFT). This can be explained by the fact that this transition involves orthogonal (non-overlapping) 
$\pi$ and $\pi^\star$ orbitals, which essentially ressembles a $\npi$ transition in terms of electron density reorganization.
Again, this conclusion fits the one of Ref.~\citenum{Gro22}.

\subsubsection{$G_0W_0$ \emph{vs} ev$GW$}

\begin{figure}[htp]
  \includegraphics[width=\linewidth]{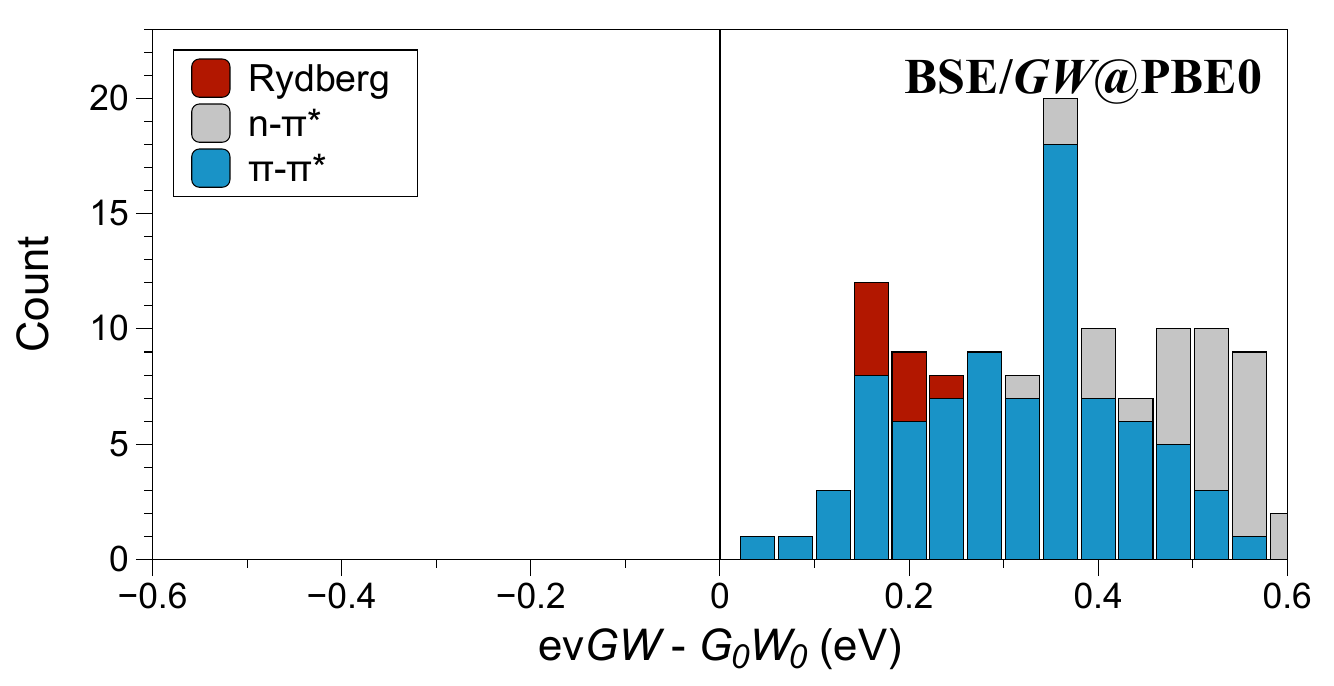}
  \includegraphics[width=\linewidth]{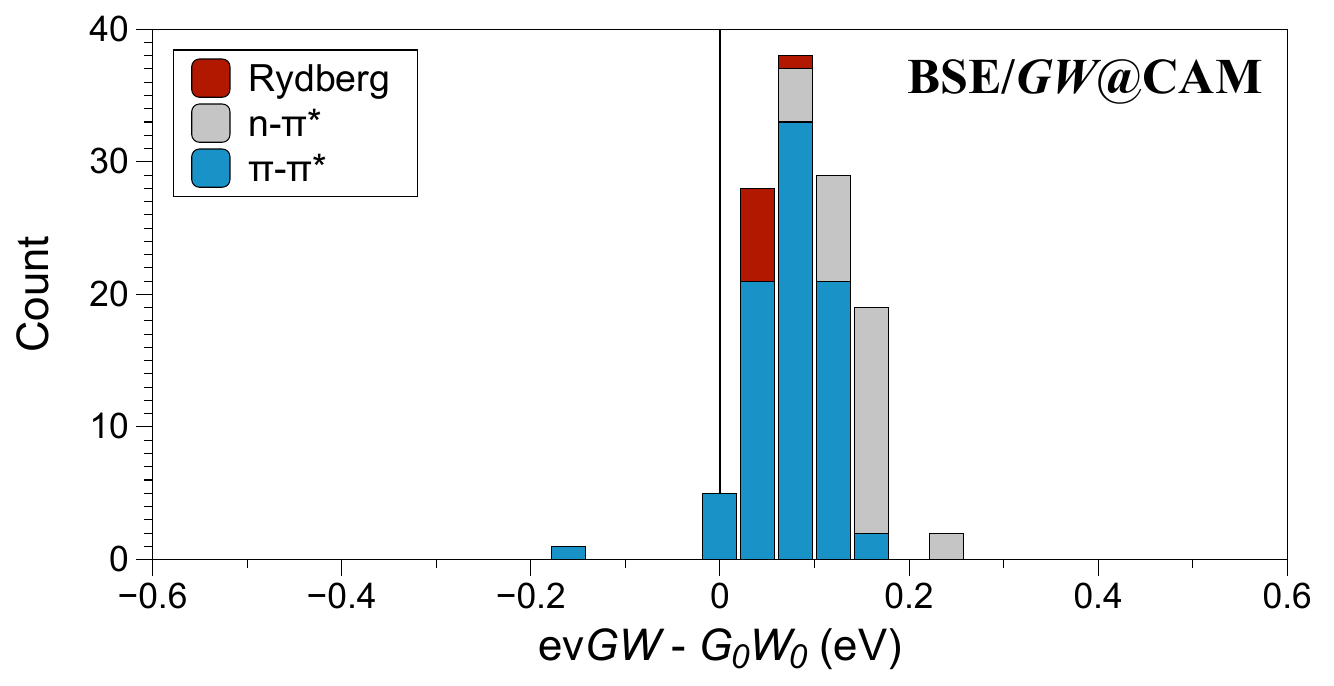}
  \includegraphics[width=\linewidth]{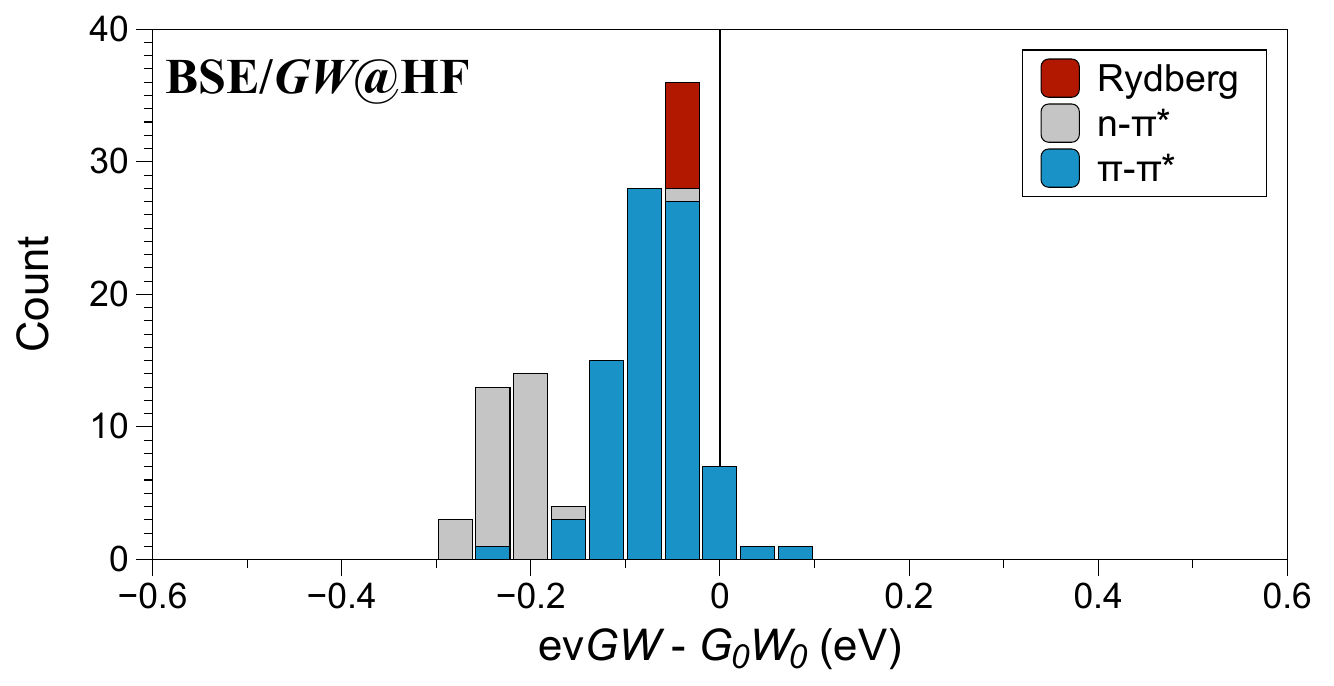}
  \caption{Histograms of the difference between (full) BSE/$GW$ results obtained on the basis of ev$GW$ and $G_0W_0$ quasiparticle energies for three different starting points. }
  \label{Fig-5}
\end{figure}

To perform a BSE calculation, one must start by computing the $GW$ quasiparticle energies and the static limit of the screened Coulomb potential. However, there are many ways of performing $GW$ calculations. \cite{Bru13,Bru15,Kni16a,Ran16c,Kap16,Car16a,Mar23a} The impact on the BSE VTEs has been previously assessed 
for other sets of compounds. \cite{Jac15a,Ran17,Gui18b} Here, we compare the two most straightforward and computationally affordable schemes, namely, the one-shot $G_0W_0$ and the partially self-consistent
ev$GW$ schemes. Histograms of the differences between the two methods can be found in Figure \ref{Fig-5} whereas additional data are given in Table S26 and Figure {S13} of the SI.

Obviously, the starting KS (or HF) quantities significantly impact the differences between the ev$GW$ and $G_0W_0$ results, hence the BSE VTEs. It is well-known that HF overlocalizes the orbitals
in $\pi$-conjugated compounds, which results in overestimated fundamental gaps. This error cannot be fully corrected by the perturbative $G_0W_0$ scheme. This is why the BSE/ev$GW$@HF VTEs are 
logically lower than their BSE/$G_0W_0$@HF counterparts by $-0.10$ eV on average. The correlation between the error on the HOMO-LUMO gap at the $GW$ level, and the error on the resulting BSE 
neutral singlet excitation energies, was clearly documented in Ref. \cite{Bru15}.  In contrast, ev$GW$ induces a large increase of the BSE excitation energies when starting from PBE0, by +0.36 eV on average. 
This hints that the PBE0 gaps are too small, well corrected at the ev$GW$ level, while still somehow underestimated at the non-self-consistent $G_0W_0$@PBE0 level. Finally, we also find a negative 
correction, albeit significantly smaller (+0.09 eV), when initiating the BSE/ev$GW$ calculations with CAM-B3LYP orbitals and energies.

From Figure \ref{Fig-5}, it is also clear that as for the gauge effects described in Section \ref{sec:gauge}, the impact of ev$GW$ depends on the nature of the considered transition: it is (relatively) mild for 
Rydberg transitions, moderate for $\ppi$ transitions and large for the $\npi$ ES. In contrast, Figure {S13} demonstrates that the differences between the two $GW$ procedures are almost independent of 
the application  of the TDA.

\subsubsection{Effect of the TDA}

\begin{figure*}[htp]
  \includegraphics[width=0.48\linewidth]{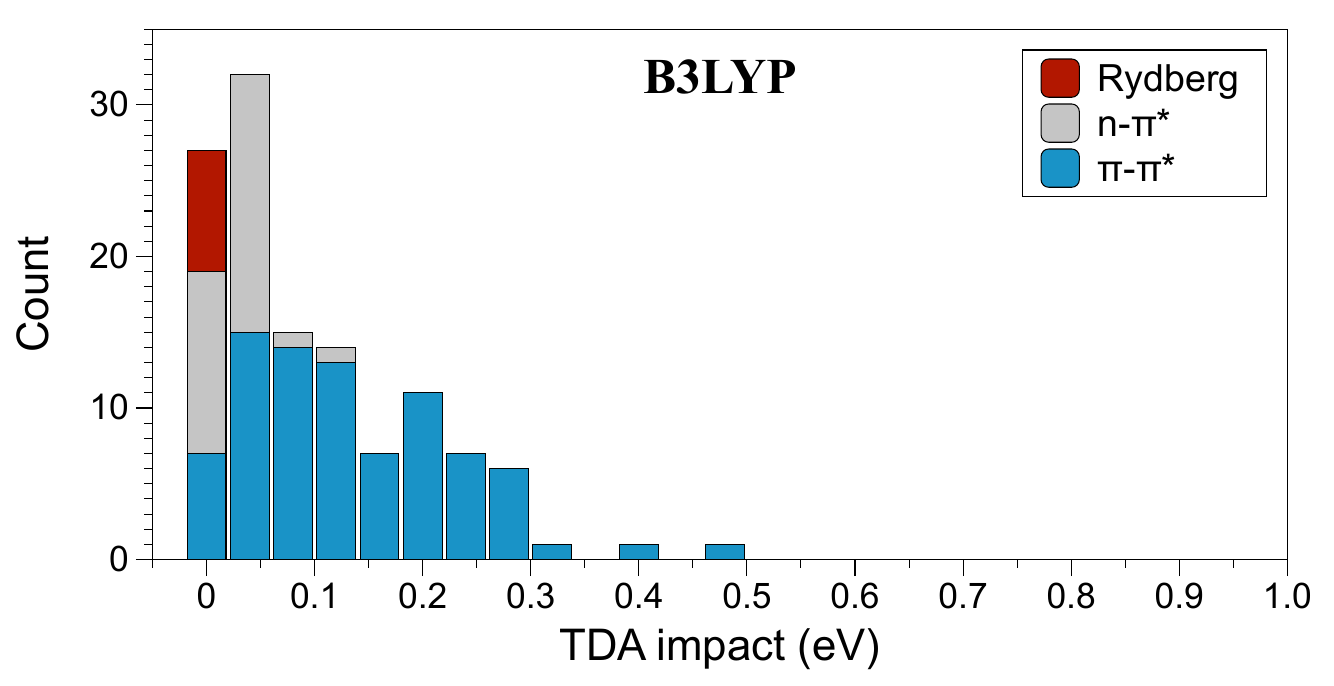}
  \includegraphics[width=0.48\linewidth]{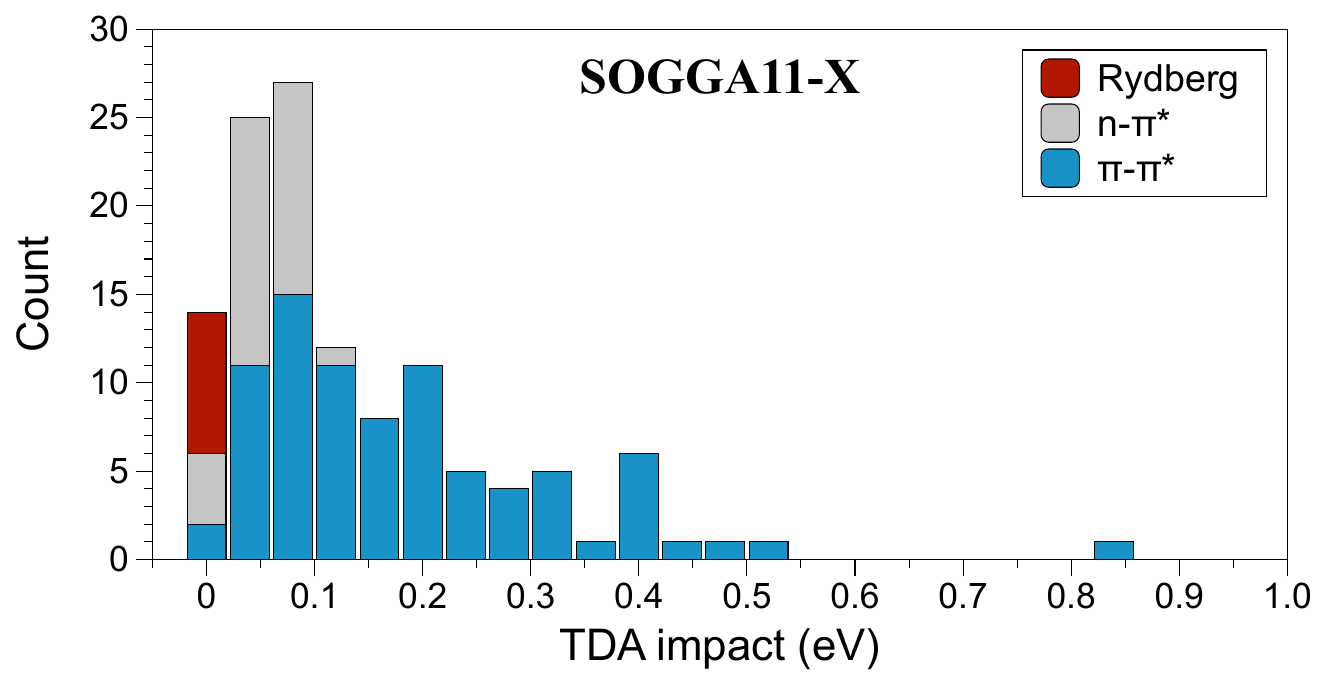}
  \includegraphics[width=0.48\linewidth]{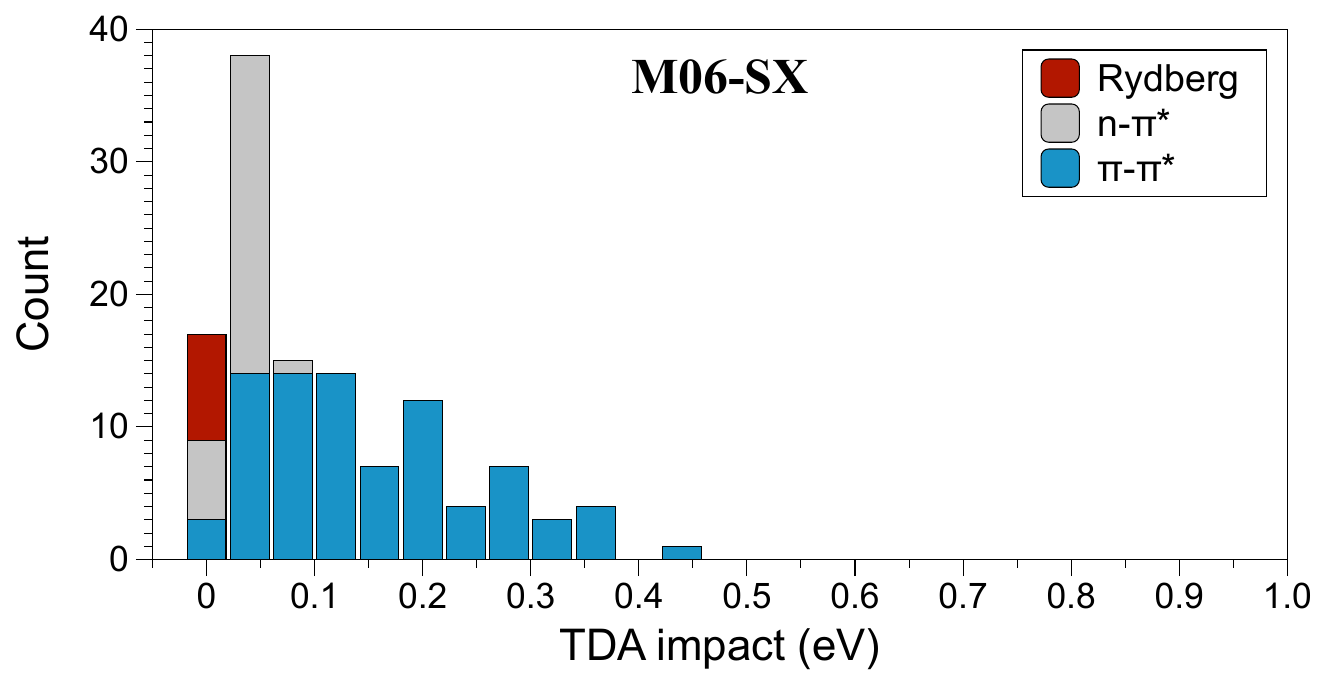}
  \includegraphics[width=0.48\linewidth]{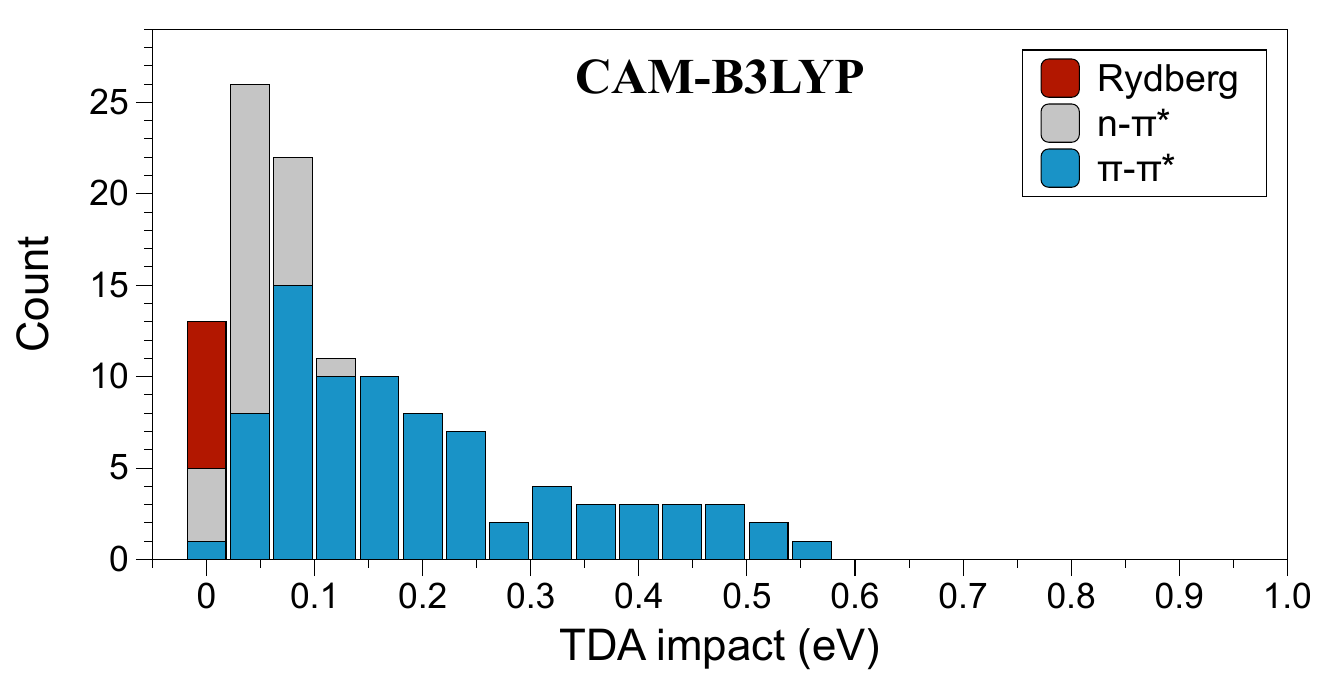}
  \includegraphics[width=0.48\linewidth]{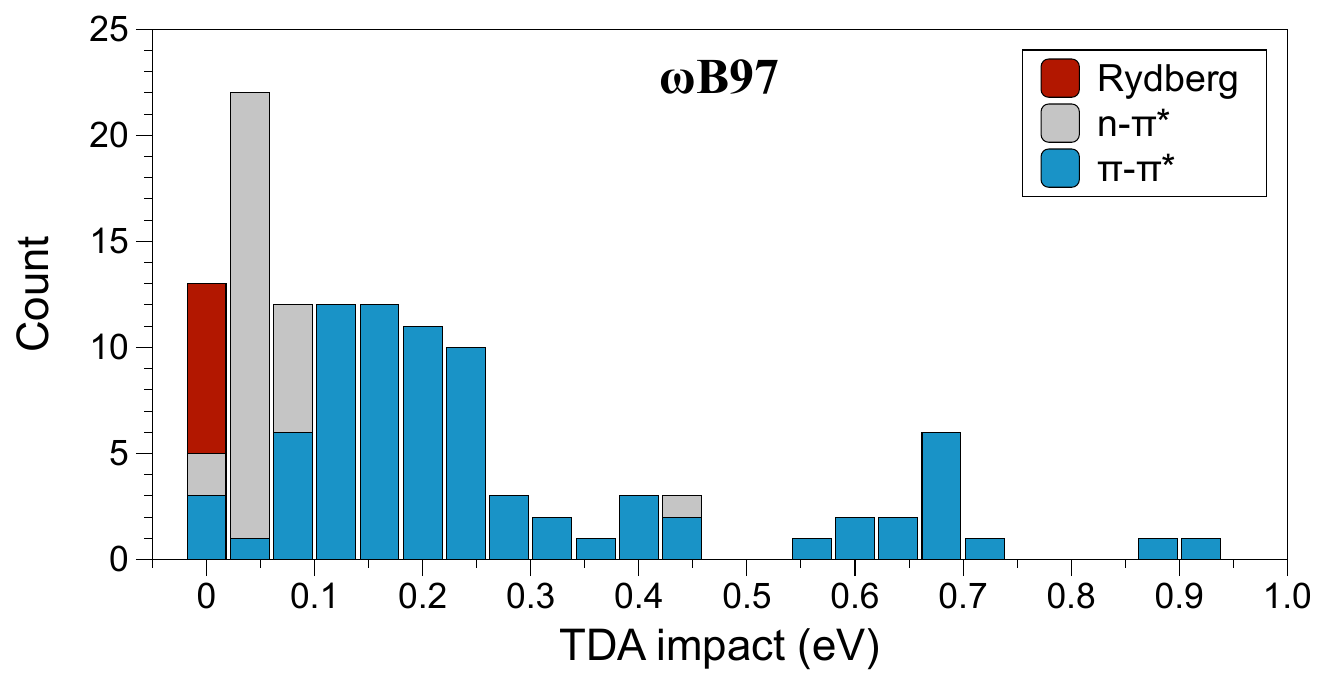}
  \includegraphics[width=0.48\linewidth]{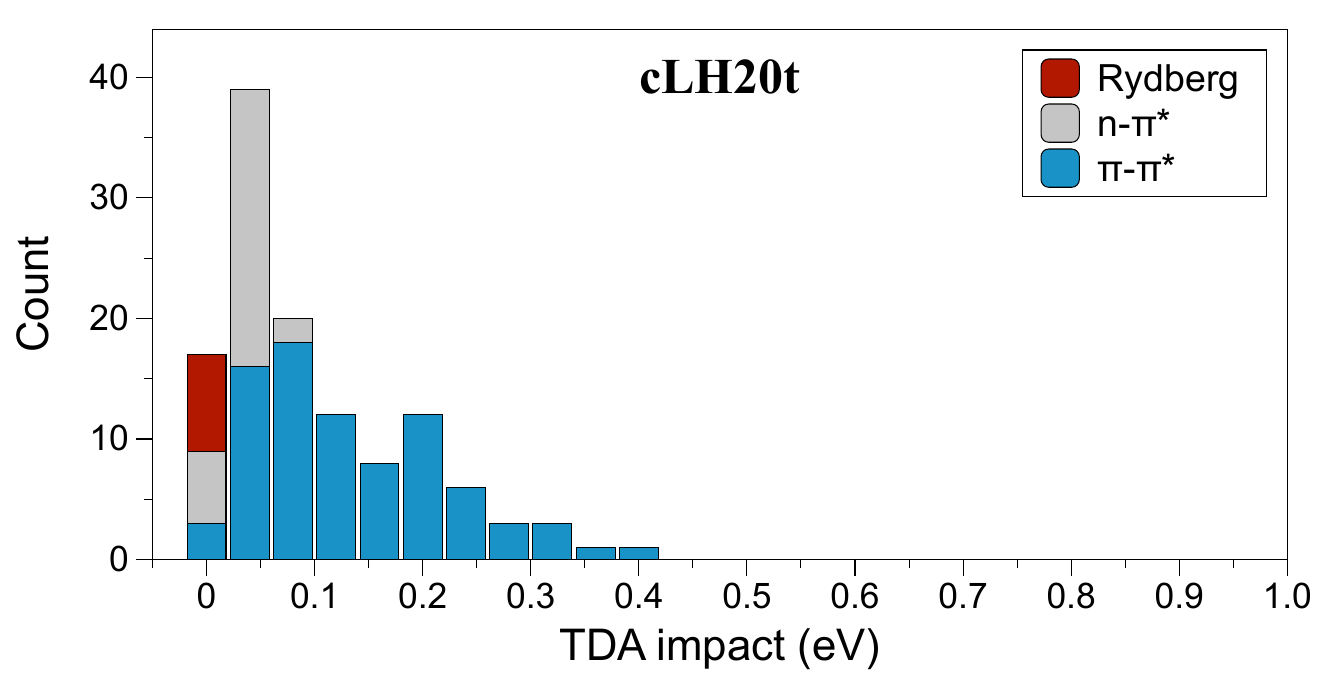}
  \includegraphics[width=0.48\linewidth]{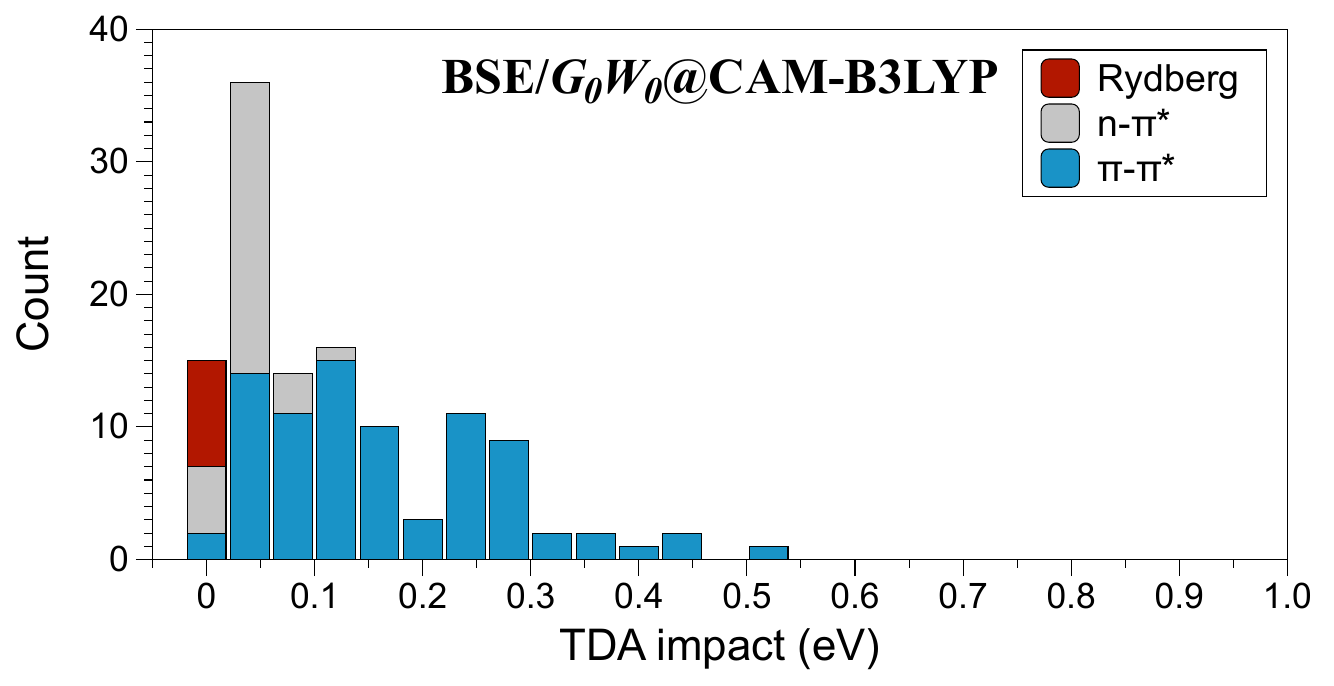}
  \includegraphics[width=0.48\linewidth]{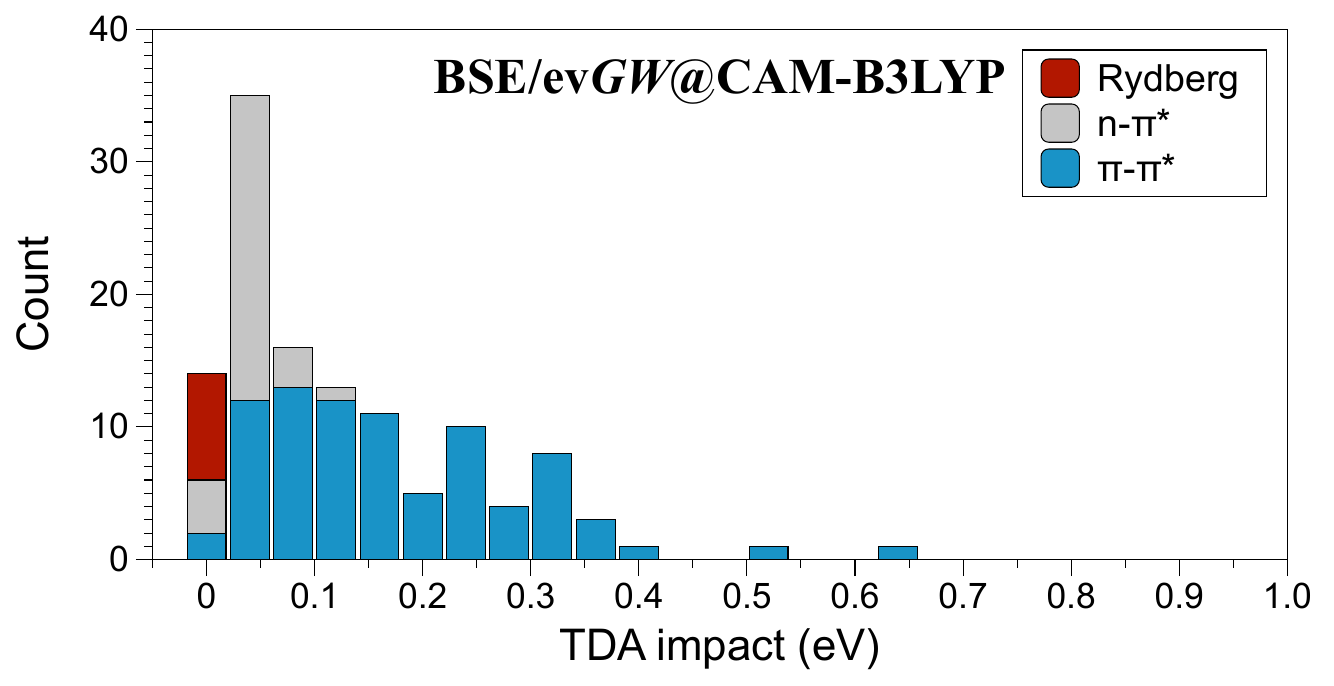}
  \caption{Impact of the application of the TDA on the VTEs (in eV) for selected TD-DFT and BSE/$GW$ schemes. Note the difference in scaling of the vertical axes.}
  \label{Fig-6}
\end{figure*}

As expected, \cite{Hir99d,Pea11,Pea12,Cha13c,Len15,Ran17,Jac17a,Lia22} applying the TDA induces an almost systematic increase in the transition energies, a statement holding true for both TD-DFT and BSE/$GW$. 
The few exceptions are related to states characterized by a significant orbital mixing with an admixture modified by the application of the TDA. $\omega$B97X-2 seems particularly prone to deliver negative TDA 
corrections.  We present histograms of the TDA corrections for B3LYP, SOGGA-11X, M06-SX, CAM-B3LYP, $\omega$B97, cLH20t, as well as two BSE/$GW$ schemes in Figure \ref{Fig-6}.
The graphs for the other XCFs as well as Tables of statistical data can be found in {Sections S5.2 and S6.2} of the SI.

\begin{figure*}[htp]
  \includegraphics[width=0.9\linewidth]{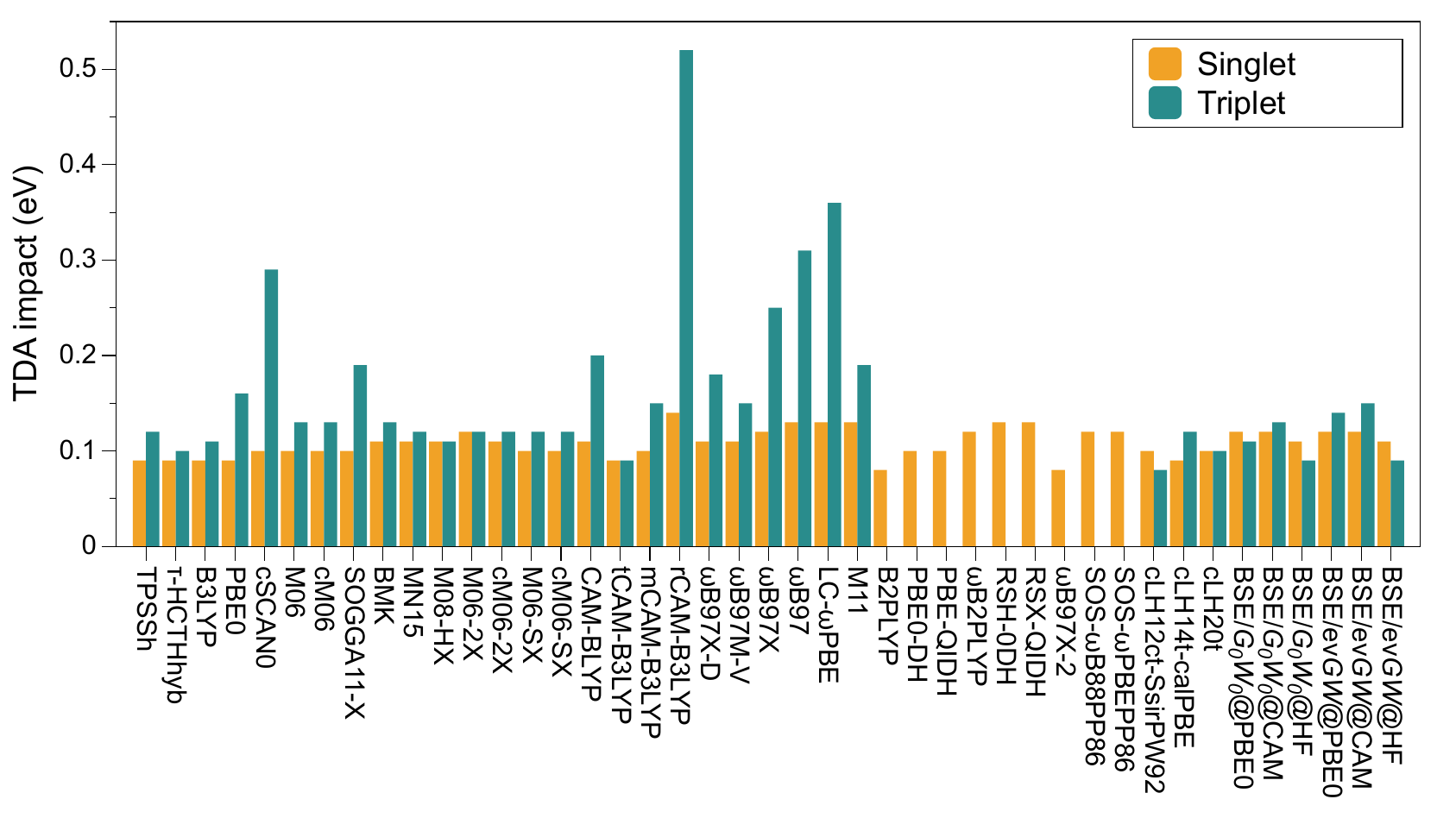}
  \caption{Average impact of the application of the TDA (in eV) for singlet and triplet ESs for various computational schemes based on TD-DFT and BSE/$GW$.}
  \label{Fig-7}
\end{figure*}

For all XCFs and for both TD-DFT and BSE, the impact of the TDA is negligible for Rydberg ESs (typically less than 0.01 eV), rather small for $\npi$ transitions (ca.~0.05 eV), and significant for $\ppi$ excitations 
(from 0.1 to 0.3 eV depending on the functional). The relative ranking of the ESs can therefore be affected. As expected as well, for the vast majority of the tested XCFs, the 
upshifts induced by the TDA are larger for the triplet than for the singlet transitions (Figure \ref{Fig-7}). For TD-DFT, we could find only one clear exception to this general rule, namely cLH12ct-SsirPW92, 
whereas the average variations are essentially equivalent for ESs of both spin symmetries with both M06-2X and M08-HX.  Consistent with an earlier work, \cite{Jac17a} the picture is more contrasted in BSE/$GW$ with 3-out-of-6 
schemes leading to larger corrections for singlet than for triplet transitions (Figure \ref{Fig-7} and Table S27).  In TD-DFT, the average upshifts for the singlets are +0.09/+0.10 eV for the least TDA-sensitive XCFs, that is,
the GHs including a small fraction of exact exchange (EXX), the LHs, and several DHs (B2PLYP, PBE0-DH, PBE-QIDH, and $\omega$97X-2).  The impact of the TDA on the singlet ESs becomes slightly larger for 
RSHs and some DHs, e.g., +0.13 eV for M11, $\omega$B97, and RSH-0DH.  Nevertheless, the changes are rather systematic and also quite insensitive to the selected functional: for a given molecule, the ESs 
that are strongly affected by the TDA are the same irrespective of the chosen XCF. The same holds for BSE/$GW$ with an average increase of the singlet energies of +0.11/+0.12 eV depending on the selected variant.   
As expected, \cite{Lia22} the picture is significantly different for the triplet ESs in TD-DFT, with corrections clearly increasing with the amount of EXX and average energetic shifts becoming very large 
for XCFs, such as LC-$\omega$PBE (+0.36 eV) and rCAM-B3LYP (+0.52 eV).  SCAN0 is an outlier with an average increase of +0.29 eV, despite a rather small fraction of EXX (25\%). The peculiar behavior of SCAN0
was reported earlier. \cite{Lia22} Consistent with an earlier observation,  \cite{Gro21} the LHs seems relatively unaffected by the use of the TDA. Yet, the increase of triplet VTEs,  ranging from  +0.09 eV to +0.12 eV for the three 
tested LHs, is rather similar to the one obtained with B3LYP, namely +0.11 eV.  For the triplet ESs, the increase in the BSE VTEs seems smaller with $G_0W_0$ than ev$GW$ and smaller when starting from HF than KS (PBE0 and 
CAM-B3LYP) eigentstates. Nevertheless, the impact of the TDA on the triplet transition energies appears slightly smaller with BSE@CAM-B3LYP than with TD-DFT@CAM-B3LYP (as evidenced by the comparison of Tables S23 and S26).

At this stage, we believe that it is not possible to conclude if enforcing the TDA is beneficial in general. Therefore, we benchmark both flavors in the following.

\subsubsection{TD-DFT \textit{vs} TBE}
\label{sec:comp_tbe}

\begin{figure*}[htp]
  \includegraphics[width=0.9\linewidth]{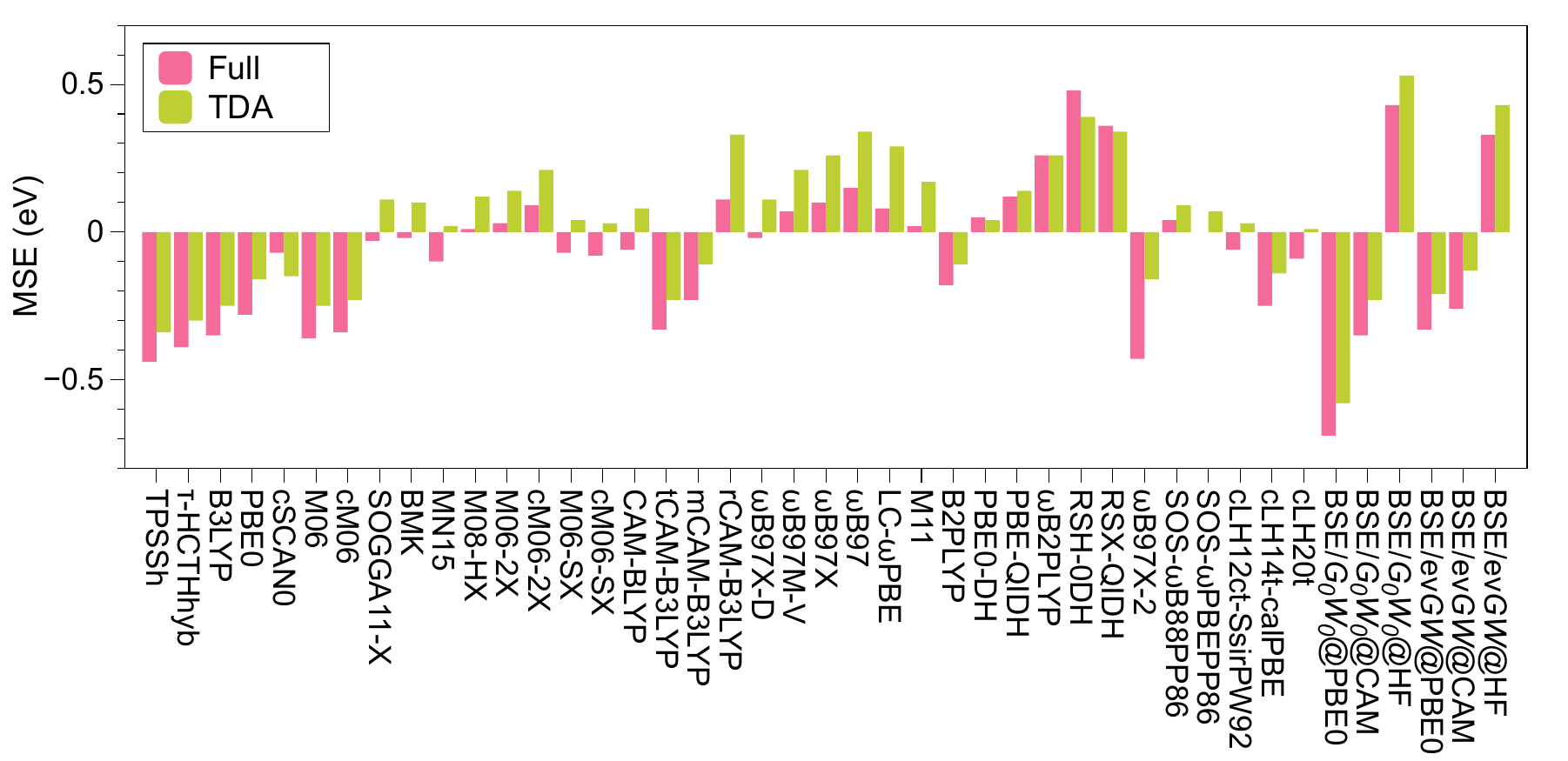}
  \includegraphics[width=0.9\linewidth]{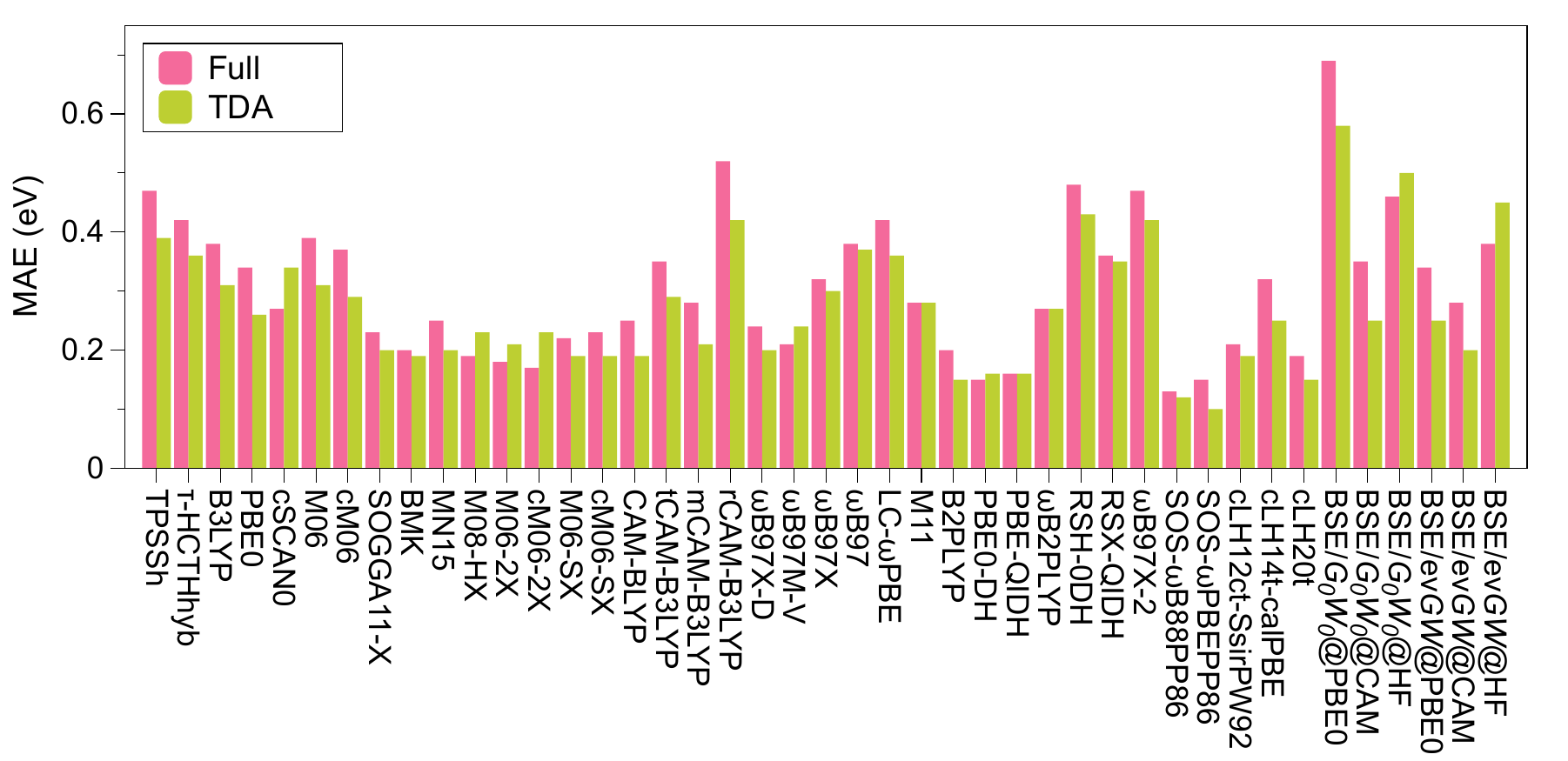}
  \includegraphics[width=0.9\linewidth]{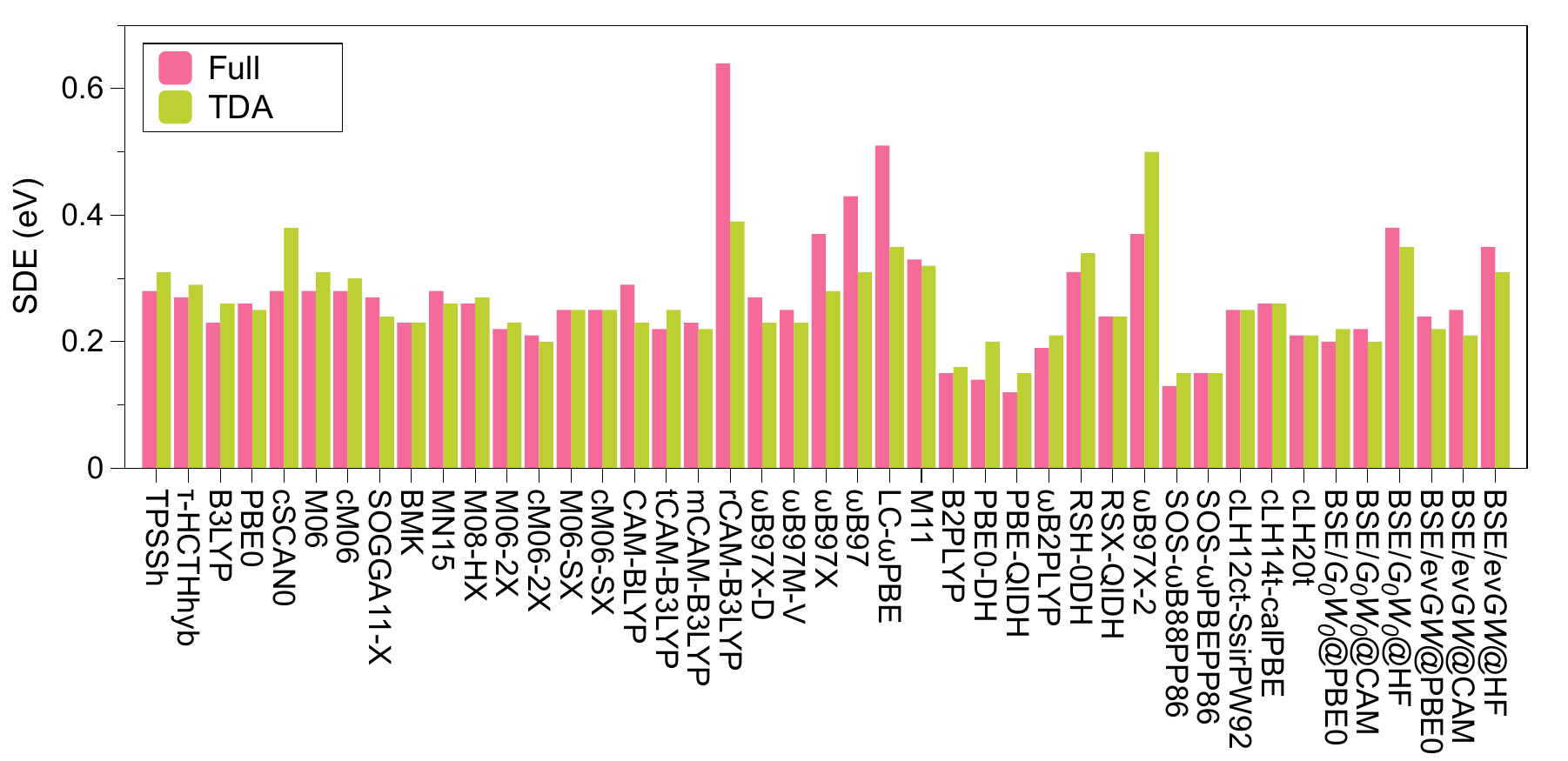}
  \caption{From top to bottom: MSEs, MAEs, and SDEs (in eV) for all XCFs using both the full (pink) and the TDA-based (green) approaches for all evaluated TD-DFT and BSE/$GW$ models.}
  \label{Fig-8}
\end{figure*}

We now turn towards the assessment of all XCFs against the TBEs. Full data can be found in Table S24-S25 (for TD-DFT) and {S29} (for BSE) in the SI, together with histograms of the errors, whereas a graphical 
representation of the MSEs, MAEs, and SDEs can be found in Figure \ref{Fig-8}. We start our analysis below by concentrating on the TD-DFT data before turning our attention to the BSE/$GW$ results.

As expected, for the GHs, the evolution of the TD-DFT MSEs (roughly) follows the percentage of EXX: \cite{Lau13} it is negative for XCFs with low EXX percentage and becomes close to zero from 40\%\ of EXX. 
The impact of applying the TDA and thus upshifting all values is also clear. For example, it decreases the MSE of PBE0 from $-0.28$ to $-0.16$ eV while increasing its M06-2X counterpart from $+0.03$ to $+0.14$ eV.
Within the TDA, all RSHs have positive MSEs, except for tCAM-B3LYP and mCAM-B3LYP, whereas in TD-DFT, $\omega$B97X-D shows a tiny average deviation of $-0.02$ eV. Two of the tested LHs, namely, 
cLH12ct-SsirPW92 and cLH20t, also show small negative (positive) MSEs with TD-DFT (TDA-DFT). Amongst the tested DHs, three of them (PBE0-DH, SOS-$\omega$B88PP86, and SOS-$\omega$PBEPP86) 
show absolute MSEs smaller than 0.10 eV, irrespective of the application of the TDA.

Let us now turn to the MAEs. The MAEs for each ES subsets can be found in Tables S24 and S25 of the SI. For the most accurate XCFs, considering only transitions with a dominant single-excitation character 
($\%T_1 \geq 85\%$)  slightly decreases the MAEs, which is the expected trend since adiabatic TD-DFT is not equipped to describe properly ESs having a partial double-excitation character. In contrast, the relative 
deviations obtained for $\ppi$, $\npi$, and Rydberg ESs is strongly dependent on the XCF, with a magnified XCF-sensitivity for the latter type of states. \cite{Lia22} Amongst the GHs, the smallest MAEs are obtained 
for TD-cM06-2X and TD-M06-2X, the former functional delivering an error as small as 0.14 eV for the transitions showing $\%T_1 \geq 85\%$, a rather remarkable result. It is noteworthy that, with these two XCFs, the 
final error is driven by the singlet  (and not triplet) ESs. Comparisons for smaller molecules using Thiel's set have indeed highlithed TD-M06-2X as a valuable method for singlet-triplet transitions. \cite{Jac10c} When 
applying the TDA, the smallest GH MAE is obtained with BMK: 0.19 eV or 0.16 eV for the $\%T_1 \geq 85\%$ transitions. Amongst the RSHs, the smallest MAEs are 0.21 eV for TD-$\omega$B97M-V 
and 0.19 eV with TDA-(c)M06-SX and TDA-CAM-B3LYP. Globally, the long-range corrected XCFs with large attenuation parameters ($\omega$B97 and LC-$\omega$PBE) are significantly less satisfying,
but the present set does not contain clear CT ESs that might require such XCFs. \cite{Loo21a} Amongst the three tested LHs, cLH20t appears to be the most accurate for the present dataset with MAEs of 0.19 eV (TD-DFT) 
and 0.15 eV (TDA-DFT). The accuracy of TD-DFT/TDA-DFT can be also improved by using the more computationally demanding DHs with, e.g., MAEs of approximately 0.15 eV for PBE0-DH and PBE-QIDH. Even smaller 
values are obtained with XCFs designed to reproduce transition energies, namely SOS-$\omega$B88PP86 and SOS-$\omega$PBEPP86, the later providing a quite astonishing MAE of 0.10 eV when combined with the TDA. 
However, relatively poor estimates are obtained for the few Rydberg transitions considered here. 

An important feature is consistency and this is measured by the SDE which assesses the systematic character of the errors. Gratifyingly, in both TD-DFT and TDA-DFT, the smallest SDEs are obtained with cM06-2X in the GH series 
and cLH20t in the LH series: the MAE and SDE performances parallel each other for these two groups of XCFs. This is not the case for the RSHs at the TD-DFT level, since tCAM-B3LYP, which was tuned to reproduce transition 
energies in diarylethene photochromes, delivers a small SDE of 0.22 eV, despite a rather large MAE of 0.35 eV. Within the TDA, the smallest (RSH) SDEs are reached with mCAM-B3LYP, CAM-B3LYP and $\omega$B97X-D.  
In the DH series, PBE-QIDH provides a SDE of 0.12 eV at the TD-DFT level (a set that does not contain triplets, however). When enforcing the TDA, a SDE of 0.15 eV can be obtained with the same functional, as well as 
with the already mentioned SOS-$\omega$B88PP86 and SOS-$\omega$PBEPP86 XCFs.

In short, it appears possible to select GHs, RSHs, and LHs providing an accuracy around 0.15--0.20 eV, the best performing schemes being likely TD-cM06-2X and TDA-cLH20t. To reach more accurate estimates, DHs are 
relevant alternatives, with MAEs and SDEs not exceeding 0.15 eV with both SOS-$\omega$B88PP86 and SOS-$\omega$PBEPP86. The performances of these two XCFs are clearly superior to the one obtained with 
CIS(D), and rather similar to the one of ADC(2) and CC2. Given this outcome, it is quite evident that benchmarking DH XCFs using ``reference'' data produced with these second-order methods is unfounded (\emph{vide infra}).

Compared to the recent results of Refs.~\citenum{Lia22} and \citenum{Cas21b}, where the original QUEST data were used to benchmark XCFs within the TDA, one notices that the conclusions regarding 
the most accurate XCFs are quite similar, with, on the one hand, valuable performances of BMK, M06-SX, $\omega$B97X-D, SOGGA11-X, and CAM-B3LYP (within the TDA) \cite{Lia22} and, on the other hand, 
excellent results obtained with PBE-QIDH, SOS-$\omega$B88PP86 and SOS-$\omega$PBEPP86 in the DH category. \cite{Cas21b} The order of magnitude of the errors reported in these different contributions 
is also reasonably consistent with the current ones, which is reassuring and hints that the conclusions are rather robust (for organic compounds). In an older assessment relying on Thiel's set, \cite{Jac09c} the 
conclusions differed with, for instance, PBE0 providing a MAE significantly smaller than BMK (0.24 \emph{vs} 0.34 eV as compared to 0.34 \emph{vs} 0.20 eV here). Likewise, for a given XCF, 
earlier benchmarks have typically delivered larger MAEs than the ones reported here. \cite{Jac09c,Goe09a,Car10,Ise12,Lea12}  For instance, Ref.~\citenum{Car10}, which relied on experimental references,
reported a MAE of 0.55 eV for TD-PBE0 as compared to 0.34 eV here. Likewise, Gordon's analysis, which also considers experimental reference energies, reported a MAE of 0.59 eV for CAM-B3LYP, as compared to 
0.25 eV in the present case. \cite{Lea12} This clearly indicates that the quality of the reference energies is crucial for providing a reliable assessment of these methods. We here recall that comparisons between 
theoretical VTEs and experimental values are inherently biased. Nevertheless, previous benchmark studies already highlighted BMK \cite{Goe09a} and M06-2X \cite{Jac10c,Ise12,Lea12} as adequate XCFs for 
ES calculations within the TD-DFT and/or TDA-DFT formalisms.

\subsubsection{BSE/$GW$ \textit{vs} TBE}

Let us now discuss BSE/$GW$. The plots displayed in Figure \ref{Fig-8} and the data of Table S29 indicate that HF starting quantities are not recommended: they lead to strong overestimations of the VTEs with 
MSEs larger than 0.3 eV irrespective of the applications of the TDA and/or the selection of ev$GW$ or $G_0W_0$. Large errors are obtained for all subsets of ESs.  Although they lead to negative MSEs, KS 
orbitals are clearly more adequate for the present systems. Globally, the more rudimental $G_0W_0$ scheme is detrimental to the final accuracy, especially with PBE0 since the VTEs are largely underestimated, 
consistent with earlier benchmarks performed on Thiel's set. \cite{Jac15a,Bru15,Gui18b} $G_0W_0$ requires XCFs with a large share of EXX as, for example, CAM-B3LYP. \cite{Bru15}  The smallest MAE 
and SDE, both of 0.20 eV, are reached with TDA-BSE/ev$GW$@CAM-B3LYP.  Such deviations can likely be viewed as rather large since several XCFs deliver more accurate results when used in TD-DFT calculations. 

Consistent with earlier findings, \cite{Bru15,Jac17a,Ran17,Gui18b} we have found that BSE/$GW$ provides an unbalanced treatment of singlet and triplet ESs. Indeed, three BSE schemes, namely BSE/ev$GW$@CAM-B3LYP, 
TDA-BSE/ev$GW$@PBE0,  and TDA-BSE/ev$GW$@CAM-B3LYP, deliver MAEs smaller than 0.15 eV for the singlet transitions, with an average error as small as 0.11 eV for the latter approach. For comparison, the most 
effective XCFs amongst the GHs, RSHs, and LHs produce significantly larger MAEs for the same subset (0.18 eV for TD-cLH20t and 0.19 eV for TDA-cLH20t), whereas even the best DH  (0.13 eV for TDA-SOS-$\omega$PBEPP86) 
cannot outcompete this tiny MAE. For the same BSE schemes, the MAE on the triplets increases to  0.48 eV,  0.40 eV, and 0.33 eV, respectively, with a significant improvement associated with the TDA. The stability of the 
BSE/ev$GW$ data, despite the significant differences between the KS starting points, from the PBE0 GH to the CAM-B3LYP RSH, indicates the advantage of the partially self-consistent ev$GW$ scheme that dramatically limits 
the starting point dependency. This is also clearly seen in the data given in Table S28 and Figure {S15} in the SI that demonstrate that the impact of starting with CAM-B3LYP instead of PBE0 is rather small with 
BSE/ev$GW$ (+0.08 eV), but becomes significant with BSE/$G_0W_0$ (+0.34 eV). For comparison, the corresponding TD-DFT shift is +0.22 eV. In short, we confirm that BSE/ev$GW$ can be a valuable alternative to 
TD-DFT for singlet ESs (only) since the resulting errors are similar to those of ADC(2) and CC2. This conclusion is in inline with the results obtained in Ref.~\citenum{Jac15b}. Optimally-tuned RSHs \cite{Ran17}, 
fully self-consistent $GW$ schemes \cite{Gui18b} or hybrid BSE/TD-DFT approaches, \cite{Hol18b} are three possible pathways to improve the description of the triplet ESs within the BSE formalism. 

\subsection{Further discussion}

\begin{figure*}[htp]
  \includegraphics[width=1.0\linewidth]{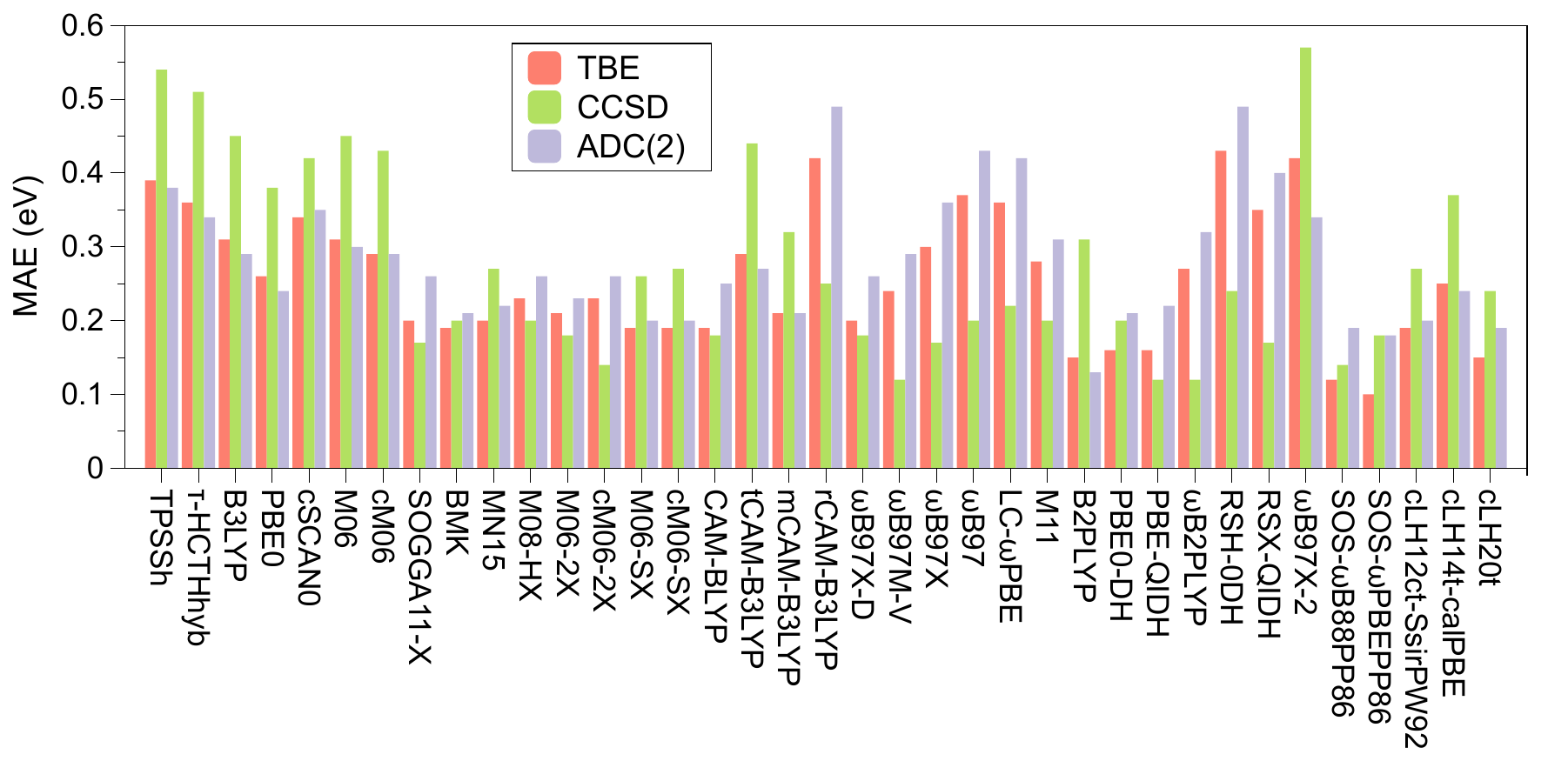}
  \caption{MAEs (in eV) obtained at the TDA-DFT level using the TBE (red), CCSD (green), and ADC(2) (purple) data as reference for various XCFs.}
  \label{Fig-9}
\end{figure*}

Another aspect that we already briefly evoked in Section \ref{sec:comp_tbe} is the importance of accurate references for benchmarking purposes. This central point is illustrated in Figure \ref{Fig-9} where we provide the 
MAEs obtained at the TDA-DFT level considering the entire set of XCFs using the TBE, CCSD, or ADC(2) values as references. We underline that all ESs are present for all methods (both reference and tested approaches), 
allowing well-grounded comparisons. Obviously, benchmarking TDA-DFT with CCSD references is not a fantastic idea: the MAE of the GHs with a low EXX percentage significantly increases (e.g., B3LYP's MAE goes from 
0.31 to 0.45 eV), whereas the MAE of RSHs with a large  attenuation parameter is strongly underestimated (e.g., $\omega$B97X's MAE goes from 0.30 to 0.17 eV). This is the logical consequence of the tendency of 
CCSD to overestimate the VTEs, as illustrated in Figure \ref{Fig-2}. In other words, CCSD favors XCFs with a large amount of EXX, since they exhibit the same overshooting predisposition.  

From the wave function results of Table \ref{Table-2}, one could argue that opting for ADC(2) as a reference method is a better choice since it presents a more balanced error pattern with a near-zero MSE of $-0.01$ eV 
(Table \ref{Table-2}). ADC(2) is actually the reference method of the very recent DELFI ``oracle'' whose aim is to predict the most suitable XCF for a given compound through a machine-learning-based approach. \cite{Ava24} 
From Figure \ref{Fig-9}, one indeed notices quite similar MAE patterns for most XCFs when using the TBE or ADC(2) values as references. However, by taking a closer look, one sees that the MAEs predicted for the 
most accurate XCFs are significantly overshot when one relies on ADC(2). For example, SOS-$\omega$B88PP86's and SOS-$\omega$PBEPP86's  MAEs are shifted from 0.12 to 0.19 eV and from 0.10 to 0.18 eV, 
respectively when replacing the TBEs by the ADC(2) data. Actually, ADC(2) would also incorrectly foresee B2PLYP as the best performing DH with a MAE of 0.13 eV. In other words, ADC(2) can probably be employed to 
pinpoint poor functionals, but appears not accurate enough to highlight the best ones.

\begin{figure*}[htp]
  \includegraphics[width=0.9\linewidth]{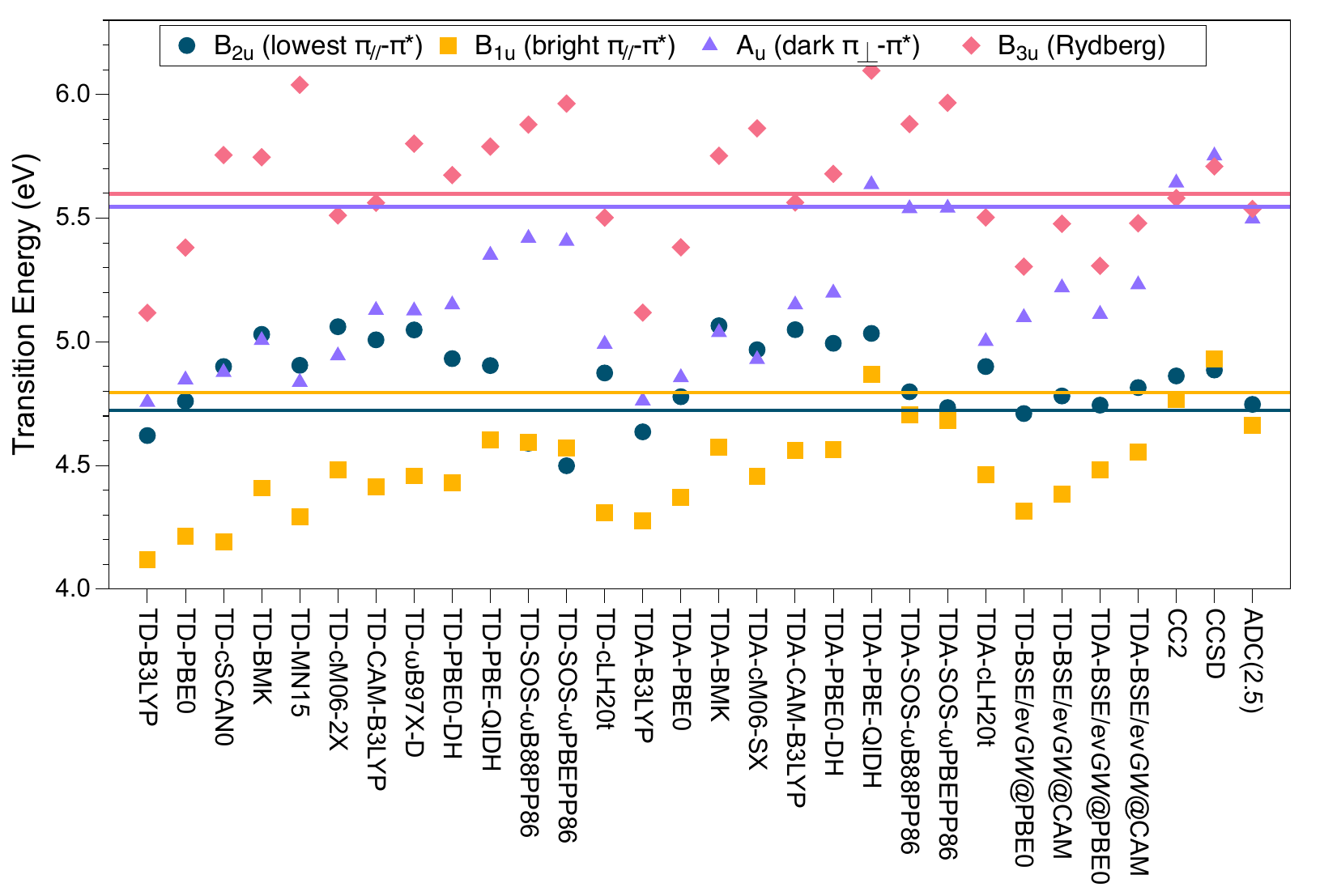}
  \caption{Excitation energies of the four lowest \emph{ungerade} singlet ESs of tolan computed with various computational schemes. The horizontal lines correspond to the TBE associated with each state as reported in Table \ref{Table-1}. }
  \label{Fig-10}
\end{figure*}

As performing an exhaustive benchmark study for a long list of ESs is a rather tedious task (especially in terms of state identification), it might also be useful to have a quick challenging test when, for example,
developing new methods or testing new protocols. In Figure \ref{Fig-10}, we provide, for a representative set of methods, the energies of the lowest four \emph{ungerade} 
singlet ESs of tolan. They encompass the lowest-lying and nearly dark $\ppi$ state ($B_{2u}$), a very bright $\ppi$ state ($B_{1u}$), a Rydberg state ($B_{3u}$), and the already mentioned state 
involving a transition between orthogonal $\pi_\perp$ and $\pi^\star$ orbitals ($A_u$). In short, these four states represent a quite diverse and somehow representative panel. Interestingly, the first two and  
last two ESs are (nearly) energetically degenerate according to the data of Table \ref{Table-1}.  As can be seen in  Figure \ref{Fig-10}, while the three reported wave function approaches reasonably reproduce the energy 
pattern (albeit with some state inversions), TD-DFT and TDA-DFT calculations based on GHs, RSHs, and LHs provide unsatisfactory results: the bright ES is typically over stabilized, the $B_{2u}$ and 
$A_u$ ESs are too close in energy, while the Rydberg ES is somehow randomly located. BSE/$GW$ also over stabilizes the $B_{1u}$ states but it more correctly separates the $B_{2u}$ and 
$A_u$ ESs. The DHs also struggle in locating the Rydberg and $\pi_\perp\pi^\star$ states, though some of them are more successful for the three valence states.

\section{Conclusions}
\label{sec:CCl}

In an effort to extend our database of accurate VTEs to larger molecules than those in the original QUEST database, and to address compounds more typical of those modeled in TD-DFT applications, we have 
determined 122 VTEs in 13 organic chromophores using CC3, and, when possible, CCSDT, to establish reference values. This new set maintains a reasonably good balance between singlet and triplet transitions, 
as well as between $\ppi$ and $\npi$ excitations, but includes only 8 Rydberg transitions, making it valence-biased and lacking in CT transitions. Among these 122 VTEs, 106 have a $\%T_1 \ge 85\%$, indicating 
ESs with strong single-excitation character for which CC3 is particularly well-suited, often delivering chemically accurate results for smaller compounds. Unfortunately, this statement cannot be definitively confirmed 
here, as CC4, CCSDTQ, and FCI estimates are beyond current computational capabilities. Therefore, we cannot provide a precise error bar for our TBEs. Nevertheless, it is reasonable to assume that the transition 
energies listed in Table \ref{Table-1} currently constitute the best theoretical reference values available for the compounds in Figure \ref{Fig-1}.

With these reference values at hand, we benchmarked more computationally affordable methods.  {Indeed, CC3 is generally too expensive to be used in practice. To give an order of magnitude, computing
the four CC3/AVTZ VTEs of aza-BODIPY took us 29 days of (human) time on a 2Tb/48 cores node, though such timing of course depends on the details of both the implementation and architecture. In our
benchmark,} we first considered 14 second- and third-order wave function methods. A key finding is that the predictions of CCSD are poorer than those of CC2 and ADC(2) in terms of absolute errors (which 
was expected) and standard deviations (which differs from results obtained for smaller compounds). The conclusion that CC2 is less consistent than CCSD does not hold for larger, real-life molecules. While 
some might view this finding as good news for practical applications, as CC2 is a relatively cheap method ({\emph{e.g.}, the same four ESs of aza-BODIPY  could be computed in less than 1h on a standard
128Gb node}), our data suggest that significant improvements might require much more expensive wave function approaches like CC3 or CCSDT. Correcting CCSD VTEs with perturbative estimates of the triples 
still does not outperform CC2 in terms of absolute deviation. Consistent with the results obtained in the original QUEST, we found that the ADC(2.5) protocol, which averages ADC(2) and ADC(3) VTEs, is the best 
performing method among all $\mathcal{O}(N^5)$ and  $\mathcal{O}(N^6)$ wave function methods we tested here, although it suffers from a small, nearly systematic underestimation trend. {In practice however,
the ADC(3) calculations will be the bottleneck in the ADC(2.5)  protocol.}
 
We have also considered a large set of {computationally more effective} models, including {TD-DFT with} many global, range-separated, local, and double hybrid XCF, applied with or without the TDA, 
{as well as various  BSE/$GW$ models}. These methods {typically require a fraction of the computational effort of} their wave function counterparts and are typically used when modeling organic dyes.
Gratifyingly, our conclusions {for both TD-DFT and BSE/$GW$}, both qualitatively and quantitatively, align well with the most recent benchmarks.

Among the GHs, we found that (c)M06-2X and BMK are particularly accurate when performing TD-DFT and TDA-DFT calculations, respectively. These widely available models deliver small MAEs/SDEs and are 
more accurate for triplet than for singlet transitions. Similar accuracies can be obtained with TDA-CAM-B3LYP or TD(A)-cLH20t. All these approaches provide MAEs slightly below 0.20 eV with an acceptable 
spread of errors. Further error reduction can be achieved using DHs, with PBE0-DH, PBE-QIDH, SOS-$\omega$B88PP86, and SOS-$\omega$PBEPP86 emerging as the four most accurate, especially the latter 
two, which deliver very small MAEs close to 0.10 eV for valence transitions (but not for Rydberg transitions).

For BSE/$GW$, we confirmed that partial self-consistency at the ev$GW$ level, or the proper tuning of the starting DFT functional, are necessary to achieve reasonable accuracy. The most satisfactory model, 
TDA-BSE/ev$GW$@CAM-B3LYP, delivers a smaller MAE than all TD-DFT models we evaluated, including the best DH, for singlet ESs. However, this success comes at the cost of large deviations for triplet ESs.

In conclusion, although the present database is inherently biased toward medium-sized $\pi$-conjugated organic compounds, we hope the values listed in Table \ref{Table-1} will be useful for testing emerging electronic 
structure theories for ESs resembling those in real-life quantum chemical applications. These values help bridge the gap between existing high-accuracy VTEs, obtained for more compact molecules, and 0-0 benchmarks, 
typically limited to the lowest bright ESs. We are currently: i) adding DNA bases to the QUEST database; and ii) preparing a review article containing all reference data generated since the first QUEST paper in 2018.

\section*{Acknowledgements}
This work was supported by the French Agence Nationale de la Recherche (ANR) under Contract No. ANR-20-C.E.29-0005 (BSE-Forces). I.K. received financial assistance from the state within the framework of 
the EUR LUMOMAT project and the Investissements d'Avenir program ANR-18-EURE-0012. PFL thanks the European Research Council (ERC) under the European Union's Horizon 2020 research and innovation 
programme (grant agreement no.~863481) for financial support. This research used resources of the GLiCID Computing Facility (Ligerien Group for Intensive Distributed Computing, https://doi.org/10.60487/glicid, 
Pays de la Loire, France).
\section*{Supporting Information Available}
Excited-state characterizations. Full list of raw vertical excitation energies. Additional statistical data. Geometries.

\bibliography{biblio-new}

\begin{mcitethebibliography}{207}
\bibitem[xxx()]{xxx-QC}
We used xc$\_$grid=3, cis$\_$convergence=7, scf$\_$convergence=9, thresh=14,
  and s2thresh=16.\relax
\mciteBstWouldAddEndPunctfalse
\mciteSetBstMidEndSepPunct{\mcitedefaultmidpunct}
{}{\mcitedefaultseppunct}\relax
\EndOfBibitem
\end{mcitethebibliography}

\end{document}